\documentclass[reprint,amsmath,amsfonts,amssymb,pra,aps,longbibliography,superscriptaddress]{revtex4-1}
\usepackage{graphicx}
\usepackage{siunitx}
\usepackage{braket}
\usepackage{xcolor}
\usepackage{mathtools}
\usepackage[normalem]{ulem}

\newcommand{\q}{\hat{q}}
\newcommand{\p}{\hat{p}}
\newcommand{\sqv}{\ket{0}_\text{sq}}
\newcommand{\qn}{\ket{\varnothing}_\text{GKP}}

\begin{document}

	\title{A fault-tolerant continuous-variable measurement-based quantum computation architecture}
	\author{Mikkel V. Larsen}
	\email{mikkel.vilsboell@gmail.com}
	\affiliation{Center for Macroscopic Quantum States (bigQ), Department of Physics, Technical University of Denmark, Fysikvej, 2800 Kgs. Lyngby, Denmark}
	\author{Christopher Chamberland}
	\affiliation{AWS Center for Quantum Computing, Pasadena, CA 91125, USA}\thanks{This work was done before CC and KN joined the AWS Center for Quantum Computing}
	\affiliation{IQIM, California Institute of Technology, Pasadena, CA 91125, USA}
	\author{Kyungjoo Noh}
	\affiliation{AWS Center for Quantum Computing, Pasadena, CA 91125, USA}\thanks{This work was done before CC and KN joined the AWS Center for Quantum Computing}
	\affiliation{IQIM, California Institute of Technology, Pasadena, CA 91125, USA}
	\author{Jonas S. Neergaard-Nielsen}
	\author{Ulrik L. Andersen}
	\email{ulrik.andersen@fysik.dtu.dk}
	\affiliation{Center for Macroscopic Quantum States (bigQ), Department of Physics, Technical University of Denmark, Fysikvej, 2800 Kgs. Lyngby, Denmark}

	\date{January 8, 2021}

	\begin{abstract}
		Continuous variable measurement-based quantum computation on cluster states has in recent years shown great potential for scalable, universal, and fault-tolerant quantum computation when combined with the Gottesman-Kitaev-Preskill (GKP) code and quantum error correction. However, no complete fault-tolerant architecture exists that includes everything from cluster state generation with finite squeezing to gate implementations with realistic noise and error correction. In this work, we propose a simple architecture for the preparation of a cluster state in three dimensions in which gates by gate teleportation can be efficiently implemented. To accommodate scalability, we propose architectures that allow for both spatial and temporal multiplexing, with the temporal encoded version requiring as little as two squeezed light sources. Due to its three-dimensional structure, the architecture supports topological qubit error correction, while GKP error correction is efficiently realized within the architecture by teleportation. To validate fault-tolerance, the architecture is simulated using surface-GKP codes, including noise from GKP-states as well as gate noise caused by finite squeezing in the cluster state. We find a fault-tolerant squeezing threshold of \SI{12.7}{dB} with room for further improvement.
	\end{abstract}

	\maketitle

\section{Introduction}
In measurement-based quantum computation (MBQC), gates are implemented by projective measurements on a multi-mode entangled cluster state, circumventing the complex coherent unitary dynamics required in conventional gate-based quantum computation \cite{raussendorf01}. As such, the cluster state is a critical resource for MBQC, and its number of modes and structural design defines the size of a potential measurement-induced algorithm. A particularly promising platform for scaling and controlling the structure of a cluster state is the optical continuous variable (CV) platform \cite{menicucci06,gu09}, where large cluster states can be deterministically generated and controlled, and efficiently measured by homodyne detection. This has been proven by the realizations of  large-scale CV cluster states in both one dimension \cite{yokoyama13,chen14,yoshikawa16} and two dimensions \cite{larsen19,asavanant19}. Moreover, the versatility of the CV optical platform has been further corroborated by the recent demonstrations of single- and multi-mode gates using high-efficiency projective measurements on one-dimensional \cite{asavanant20} and two-dimensional cluster states \cite{larsen20b}.

MBQC based on CV is however inherently noisy due to the impossibility of generating maximally entangled CV cluster states: The generation of maximal CV entanglement requires squeezed states of infinite squeezing and thereby infinite energy, which is not feasible. Therefore, inevitably, Gaussian noise will be added to the quantum information during computation. To combat this additive noise, information is encoded as special qubits in CV bosonic modes of infinite dimension. By encoding such qubits into the bosonic modes, using e.g. a cat-code \cite{cochrane99}, a binomial code \cite{michael16}, or the Gottesman-Kitaev-Preskill (GKP) code \cite{gottesman01}, the Gaussian noise can be corrected at the cost of being converted into Pauli errors on the encoded qubit. These Pauli qubit errors must then be corrected by some qubit quantum error correction scheme. Implementing qubit error correction efficiently in MBQC puts stringent requirements on the underlying cluster state. As an example, the local connectivity in the cluster states support only coupling between nearest-neighbor modes, so topological error correction is a natural choice for qubit error correction \cite{kitaev03}. This, in turn, requires a three-dimensional (3D) cluster state for MBQC \cite{raussendorf06,raussendorf07a,raussendorf07b}.

Different proposals on 3D cluster state generation and topological MBQC exist. Fukui \textit{et al.} \cite{fukui18} suggested a scheme for fault-tolerant MBQC based on topological error correction, but their scheme assumes the availability of a highly complex 3D cluster state of encoded qubits. Wu \textit{et al.} \cite{wu19} proposed an optical setup for the generation of a 3D cluster state using time and frequency multiplexing. However, in their proposal, gates are implemented by gate teleportation through four-mode square cluster states leading to increased gate noise. In another work of Fukui \textit{et al.} \cite{fukui20}, an all-temporally encoded 3D cluster state is proposed, but this scheme is experimentally highly challenging as it requires the construction of 12 squeezing sources and real-time feed-forward operations. Moreover, no schemes for qubit encoding and qubit error correction was put forward. The most complete work on CV MBQC to date is carried out by Bourassa \textit{et al.} in Ref. \cite{bourassa20} in which a computation architecture for the generation of a 3D cluster state combined with topological MBQC is proposed. However, the suggested architecture is based on spatial encoding, rendering the number of spatial resources very large (as this number scales linearly with the computation size). Moreover, their scheme relies on a very large number of experimentally challenging on-line swap and sum gates which they assume to be ideal. Experimental work towards topological quantum computation has been demonstrated in other platforms, including a 9-qubit code in a photonic platform with polarization encoded qubits \cite{yao12}, and a 7-qubit code in an ion-trap platform \cite{nigg14} and a superconducting platform \cite{andersen20}. Still, thousands of qubits are required for large fault-tolerant codes \cite{fowler12}.

In our work, we propose a simple, scalable, and complete architecture for topological MBQC and validate the fault-tolerance of the computation scheme. It is based on gate teleportation on parallel one-dimensional (1D) cluster states, or wires, arranged in a 3D lattice and coupled by variable beam-splitters for two-mode gates. As such, the setup is a variation of the well-demonstrated 1D cluster state generation \cite{yokoyama13,larsen19,asavanant20} with added variable beam-splitters. Combined with GKP-encoded qubits \cite{gottesman01}, the scheme allows for universal computation, while fault-tolerance is achievable by encoding logical qubits in the topological surface code \cite{bravyi98,dennis02,fowler12}. Furthermore, the scheme, being based on gate teleportation, is compatible with a recently proposed GKP correction protocol that dispenses with demanding coupling to ancillary GKP-qubits \cite{walshe20}. We validate the fault-tolerance of the full scheme by a thorough simulation that includes both noise in the GKP-qubits and---unlike previous works---gate noise caused by finite squeezing in the cluster state. As a result, when combining the topological surface code with GKP error correction in the surface-GKP code \cite{noh20}, we find a squeezing threshold of $\SI{17.3}{dB}$. We continue to propose a variation of the surface-GKP code---the surface-4-GKP code with four GKP corrections during the surface code syndrome measurements---by which we upgrade the squeezing threshold to \SI{12.7}{dB} while leaving room for further improvements. Related schemes in Refs. \cite{fukui18,bourassa20} have better thresholds, but they assume an ideal cluster state, i.e. without gate noise. We obtain a comparable threshold of $\SI{10.2}{dB}$ when ignoring gate noise.

The paper is organized as follows. In section \ref{sec:setup} we present the computation scheme and describe the implementation of the required gates. In section \ref{sec:GKP} we focus on GKP error correction within the computation scheme, and in section \ref{sec:surface} we implement the surface code for qubit error correction and validate the fault-tolerance properties by performing simulations. In section \ref{sec:conclusion} we discuss the results and conclude the paper.

\section{Computation scheme}\label{sec:setup}
\begin{figure}
	\centering
	\includegraphics[width=1\linewidth]{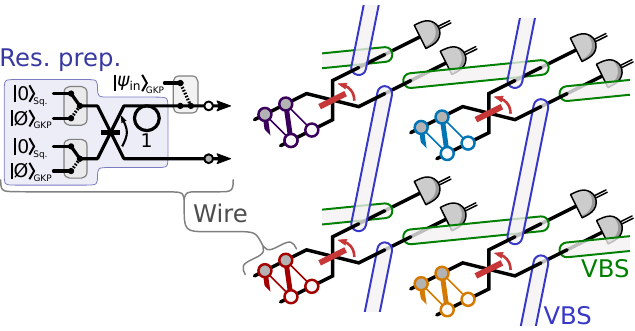}
	\caption{\label{fig:spatial} Conceptual illustration of the computation scheme: $n\times m$ wires of two-mode entangled states are prepared in $n\times m$ resource preparation gadgets (res. prep.). States with input information for computation, $\ket{\psi_\text{in}}_\text{GKP}$, are switched into each wire for computation using an optical switch. With a beam-splitter (marked with a red arrow) and two detectors for each wire, single-mode gates are implemented in each wire when feeding the resource preparation gadget with squeezed states, $\sqv$ \cite{alexander14,asavanant20}. Feeding the resource preparation gadget with special GKP states, namely qunaught states, $\qn$, allows for GKP quadrature correction of encoded GKP qubits \cite{walshe20}. Neighbouring wire setups are connected using variable beam-splitters (VBS) allowing tunable coupling for implementing two-mode gates and thereby enabling multi-mode computation.}
\end{figure}
\begin{figure*}
	\includegraphics[width=0.85\textwidth]{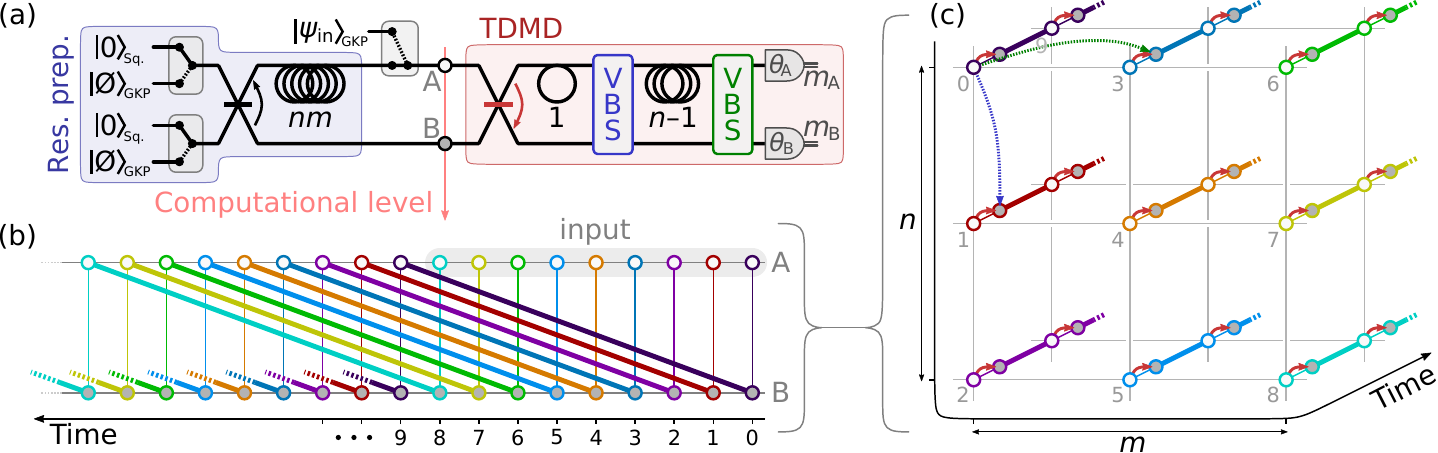}
	\caption{\label{fig:setup} (a) Temporally encoded computational setup of our scheme. The scheme consists of three parts: The resource preparation gadget (res. prep.); the computational level where the computation takes place; and the temporally delocalized measurement device (TDMD) for gate implementation by projective measurements. This scheme utilizes temporal multiplexing of two spatial modes, $A$ and $B$, marked in the computational level. (b) Wires of two-mode entanglement at the computational level shown in the time domain, here for the simple case of $nm=9$. Bold lines represent two-mode entanglement, while thin lines indicate the temporal overlap of $A$ and $B$. The wires begin with $nm$ input states in temporal modes 0 to 8, switched in using an optical switch in $A$ of the computational level. The colors of the wires have no physical meaning and are merely used to indicate different wires. (c) Wires rearranged into a 3D time lattice where the input is encoded onto the $n\times m$ end surface while gates are implemented by teleportation along the wires in the third dimension. Here, the red arrows represent the first beam-splitter of the TDMD, while the dotted blue and green arrows represent the variable beam-splitters of the TDMD. The first 10 temporal modes of (b) from 0 to 9 are labeled in (c).}
\end{figure*}

The concept of our computation scheme is illustrated in Fig.~\ref{fig:spatial}. The scheme consists of parallel \textit{wires}, each corresponding to temporally encoded one-dimensional cluster states \cite{yokoyama13,yoshikawa16} on which single-mode gates can be implemented by projective measurements using a beam-splitter (marked in Fig.~\ref{fig:spatial} with a red arrow) and two detectors \cite{alexander14,asavanant20}. Input states, $\ket{\psi_\text{in}}$, can be swapped into each wire for computation using an optical switch. To enable multi-mode computation, the setups of neighbouring wires are connected with variable beam-splitters (VBS), which allow for a tunable coupling of wires for implementing two-mode gates. This architecture may be implemented spatially as depicted in Fig.~\ref{fig:spatial} with a large grid of wire setups forming a 3D cluster state encoded in $(\text{space})^2\times\text{time}$. Such spatial encoding requires spatially scalable resources and may be possible with integrated photonics. As an alternative, in Fig.~\ref{fig:setup} we propose an all-temporal encoded version of the computation scheme which allows for a simple experimental implementation and easy scalability. In the following, while focusing on the temporally encoded architecture when describing the computation scheme in detail, all the presented methods, results, and conclusions also hold true for the spatially encoded architecture---even a combination of the spatial and temporal encoding architectures might lead to a similar computation scheme with identical conclusions.

The temporally encoded scheme in Fig.~\ref{fig:setup} consists of three parts: the preparation of resource states, the injection of input states at the computational level, and the measurements, enabled by a temporally delocalized measurement device (TDMD). Note, the term `computational level' refers to the location in the setup at which information is encoded and computation takes place. In some of our previous works this computational level is referred to as the `logic level' \cite{larsen20,larsen20b}. However, in this work we reserve the term `logic' for qubit error correction in section \ref{sec:surface}. As ancillary input for the resource preparation, we switch between squeezed vacuum states, $\sqv$, when implementing gates by projective measurements, and GKP \emph{qunaught} states~\cite{walshe20}, $\qn$, when performing GKP error correction. In section \ref{sec:GKP}, GKP error correction with ancillary $\qn$-states is described, while throughout this section we focus on gate implementation with ancillary $\sqv$-states.

At the resource preparation stage, the spatial modes $A$ and $B$ are initially occupied by squeezed vacuum states, $\sqv$, which are squeezed along the orthogonal quadratures $(\q-\p)/\sqrt{2}$ and $(\q+\p)/\sqrt{2}$, respectively. Here, $\q$ and $\p$ are the electric field amplitude and phase (or position and momentum) quadratures for which we use the $\hbar=1$ convention, corresponding to a vacuum variance of $1/2$. Each pair of squeezed states is then interfered on a balanced beam-splitter, leading to two-mode entanglement with $\q\p$-correlations. This is an approximate cluster state equivalent to a conventional two-mode squeezed state (with $\q\q$- and $\p\p$-correlations) that is phase-rotated by $\pi/4$ in both modes \cite{vanloock07,menicucci11a}. As a unitary operator for the balanced beam-splitter, we use $\hat{B}=e^{-i\pi(\q_i\otimes\p_j-\p_i\otimes\q_j)/4}$ with corresponding symplectic matrix
\begin{equation}\label{eq2:B}
	\textbf{B}=\frac{1}{\sqrt{2}}\begin{pmatrix}
		1 & -1 & 0 & 0\\
		1 & 1 & 0 & 0\\
		0 & 0 &1 & -1\\
		0 & 0 &1 & 1\\
	\end{pmatrix}
\end{equation}
acting on $(\q_i,\q_j,\p_i,\p_j)^T$ quadrature vectors, and represented graphically with an arrow pointing from mode $i$ to $j$. Note, in this work we prepare two-mode cluster states, however, we could as well have considered preparation of conventional two-mode squeezed states (with $\q\q$- and $\p\p$-correlations) which are equivalent to cluster states under phase-rotation that may be absorbed into the measurement bases.

After the interference at the beam splitter, the modes of $A$ are delayed by $nm$ temporal modes, leading to synchronization of the modes $A$ and $B$ of two-mode entangled states that initially are separated by $nm$ temporal modes \footnote{We assume the temporal mode duration and spacing to be equal. In practice, for a pulsed scheme, the temporal mode duration corresponds to the pulse width, while the $nm$-delay corresponds to a delay of $nm$ times the temporal pulse spacing.}. The result is $nm$ decoupled wires of two-mode entangled states, illustrated in the time domain in Fig.~\ref{fig:setup}(b) for $nm=9$. Here, each color indicates different wires with the bold lines indicating two-mode entanglement, while the thin lines indicate temporal overlap of $A$ and $B$. The $nm$ wires constitute the computational level in which computation is performed. Using an optical switch, $nm$ input modes to the computation can be switched into the computational level at $A$---optical switching has been demonstrated in a continuous variable quantum setting in \cite{larsen19b,takeda19}. In Fig.~\ref{fig:setup}(b) nine input modes have been switched into $A$ in temporal modes $0$ to $8$.

The $nm$ wires can be arranged in a 2D grid such that they form a 3D square lattice, as shown in Fig.~\ref{fig:setup}(c). The third dimension is in principle arbitrarily deep. As such, information is encoded on a surface while computation proceeds along the third dimension by teleportation using the TDMD. The TDMD consists of a balanced beam-splitter, two VBSs, two delays of 1 and $n-1$ temporal modes, and two homodyne detectors (HD) measuring $A$ and $B$ in bases $\q(\theta)=\q\cos\theta+\p\sin\theta$. The arrangement is illustrated in Fig.~\ref{fig:setup}(a). Each VBS can vary between two settings: when implementing single-mode gates, the VBSs are left `open' such that the modes $A$ and $B$ do not interfere, corresponding to $\hat{I}_A\otimes\hat{I}_B$; when implementing two-mode gates, one of the two VBSs are `enabled' to be functioning as a balanced beam-splitter with the symplectic matrix in Eq.~\eqref{eq2:B} interfering $A$ and $B$. Such a variable beam-splitter may be implemented in various ways, for instance as a Mach–Zehnder interferometer with a controllable phase in one arm, or by polarization control combined with polarization-dependent beam-splitters \cite{bonneau12,he17,takeda19}.

When the VBSs are left open, the TDMD simply implements a two-mode joint Bell measurement which enacts a single-mode gate teleportation through the two-mode entangled resource state \cite{alexander14,asavanant20}. The state in computation is teleported from temporal mode $k$ in $A$, $(A,k)$, to mode $(A,k+nm)$. In this process, depending on the HD basis settings, $\theta_{A,k}$ and $\theta_{B,k}$, the gate operation
\begin{equation}\label{eq2:signle_mode_gate}
	\hat{R}\left(\theta_+\right)\hat{S}\left(\tan\theta_-\right)\hat{R}\left(\theta_+\right)
\end{equation}
is implemented on the teleported state, where $\theta_\pm=(\pm\theta_{A,k}+\theta_{B,k})/2$ \footnote{Comparing with Ref.~\cite{alexander14}, a squeezing operator, dependent on the squeezing of the ancillary $\sqv$-states, is missing in Eq.~\eqref{eq2:signle_mode_gate}. This is because in this work the two-mode entangled states prepared in the resource preparation gadget are considered cluster-type states with edge weight 1 \cite{vanloock07}, similar to in Ref.~\cite{larsen20b}, instead of approximate cluster states in the language of Ref.~\cite{menicucci11a}.}. Here $\hat{R}(\theta)=e^{-i\theta(\q^2+\p^2)/2}$ and $\hat{S}(s)=e^{i\ln(s)(\q\p+\p\q)/2}$ are the rotation and squeezing operators. Note, $k$ labels the temporal modes at the computational level, while at the HDs, modes in $A$ are delayed by $n$ temporal modes relative to modes in $B$. All single-mode Gaussian gates can be implemented with two iterations of Eq.~\eqref{eq2:signle_mode_gate} \cite{ukai10}.

Enabling one of the two VBSs, two-mode gates can be implemented between nearest neighbours in the 3D time lattice. Two-mode gates between $(A,k)$ and $(A,k+1)$ are implemented by enabling the first VBS, while enabling the second VBS allows two-mode gates between $(A,k)$ and $(A,k+n)$. In the 3D time lattice of Fig.~\ref{fig:setup}(c), the VBSs are represented by dotted arrows. To encode the surface code described in section \ref{sec:surface}, we implement two different symmetric two-mode gates: $\hat{C}_Z(g)=e^{ig\q\otimes\q}$ and $\hat{C}_X(g)=e^{-ig\p\otimes\p}$. They are controlled-phase gates that displace one mode in $\p$ (or $\q$) by an amount $g\q$ (or $g\p$) controlled by the other mode. We note that $\hat{C}_X(g)$ does not correspond to a controlled-not gate. $\hat{C}_Z(g)$, or $\hat{C}_X(g)$, constitutes together with Eq.~\eqref{eq2:signle_mode_gate} a universal Gaussian gate set. In practice, $\hat{C}_Z(g)$ and $\hat{C}_X(g)$ cannot be implemented in a single computation step without some Fourier by-products of $\pi/2$ phase-rotations, $\hat{F}=\hat{R}(\pi/2)$. To implement the surface code in section \ref{sec:surface} with a minimum number of computation steps, we make use of 4 variations of $\hat{C}_Z(g)$ and $\hat{C}_X(g)$ with different by-products, each listed in table \ref{tab:two_mode_gates} with their required basis settings for implementation. These are with by-products of $\hat{F}\otimes\hat{F}^\dagger$ or $\hat{F}^\dagger\otimes\hat{F}$ when implemented on modes $(A,k)\otimes(A,k+j)$ where $j=1$ or $n$ depending on which VBS is enabled. When implementing the surface code, the gates are arranged such that the Fourier by-products cancel.

\begin{table}
	\begin{tabular}{l r}
		\hline\hline
		Two-mode gate & Basis setting, $(\theta_{A,k},\theta_{B,k},\theta_{A,k+j},\theta_{B,k+j})$\\\hline
		$(\hat{F}^\dagger\otimes\hat{F})\hat{C}_Z(g)\quad$ & $\left(-\arctan\frac{2}{g},0,0,\arctan\frac{2}{g}\right)$\\
		$\hat{C}_Z(g)(\hat{F}\otimes\hat{F}^\dagger)\quad$ & $\left(-\frac{\pi}{2}+\arctan\frac{2}{g},\frac{\pi}{2},-\frac{\pi}{2},\frac{\pi}{2}-\arctan\frac{2}{g}\right)$\\
		$(\hat{F}\otimes\hat{F}^\dagger)\hat{C}_X(g)\quad$ &
		$\left(-\frac{\pi}{2}+\arctan\frac{2}{g},\frac{\pi}{2},-\frac{\pi}{2},\frac{\pi}{2}-\arctan\frac{2}{g}\right)$\\
		$\hat{C}_X(g)(\hat{F}^\dagger\otimes\hat{F})\quad$ & $\left(-\arctan\frac{2}{g},0,0,\arctan\frac{2}{g}\right)$\\
		\hline\hline
	\end{tabular}
	\caption{\label{tab:two_mode_gates} Two-mode gates with input and output in modes $(A,k)\otimes(A,k+j)$ and $(A,k+nm)\otimes(A,k+nm+j)$, respectively, and their required basis settings. Here, $j=1$ when enabling the first VBS (marked by blue in Fig.~\ref{fig:setup}), and $j=n$ when enabling the second VBS (marked by green in Fig.~\ref{fig:setup}). The order of the tensor products are arranged with earlier temporal modes first. Note, as apparent from the basis settings, 	$(\hat{F}^\dagger\otimes\hat{F})\hat{C}_Z(g)=\hat{C}_X(g)(\hat{F}^\dagger\otimes\hat{F})$ and 	$\hat{C}_Z(g)(\hat{F}\otimes\hat{F}^\dagger)=\hat{C}_X(g)(\hat{F}^\dagger\otimes\hat{F})$. However, when implementing the gates in the surface code in section \ref{sec:surface}, it is useful to consider them individually, as we will have the Fourier by-products to cancel.}
\end{table}

Finally, as the resource squeezed states $\sqv$ are finitely squeezed, all gate implementations will inevitably produce excess noise which accumulates on the computational modes throughout the computation. Due to the Gaussian nature of the quadrature distribution of $\sqv$, this gate noise leads to a Gaussian convolution of the quadratures of all computational modes \cite{alexander14,larsen20,larsen20b}. Assuming the variance of the squeezed quadrature of $\sqv$ to be $\sigma^2=e^{-2r}/2$ (where $r$ is the squeezing parameter), the variance of the uncorrelated gate noise is 
\begin{equation}\label{eq:gatenoise}
	\sigma_\text{gate}^2=2\sigma^2=e^{-2r}
\end{equation}
which will be added symmetrically in each quadrature of the computational modes. In addition to gate noise, an implemented gate also results in a displacement of the computational modes depending on the projective measurement outcomes. Since the measurement outcomes are known, this displacement can be compensated for by another cancelling displacement operation. However, in practice, these ubiquitous displacement operations need not be executed directly onto the output modes; they can simply be accounted for in post-processing of the measurement outcomes. Therefore, in this work, we will ignore these displacements, while for practical implementation, one has to keep these in mind when analysing the measurement outcomes.

For a derivation of implemented gates considered in this section, together with their resulting gate noise and displacements, see appendix \ref{app:gates}.

\section{GKP quadrature correction}\label{sec:GKP}
As mentioned above, noise will be added to the computation modes at each gate implementation due to the finite amount of squeezing of the resource states. To correct for this noise and thus prevent noise accumulation, we consider a quadrature noise correction scheme that relies on bosonic qubit encoding in the infinite-dimensional Hilbert space. This noise correction scheme, however, comes with the cost of introducing qubit errors which must be subsequently corrected by a qubit error correction scheme. The first correction layer, the quadrature correction scheme, will be discussed in this section while the second correction layer, the qubit error correction, will be the subject of section \ref{sec:surface}. 

Several schemes for encoding qubits into bosonic harmonic oscillators of infinite Hilbert space dimension exist, including cat-codes \cite{cochrane99}, binomial codes \cite{michael16}, and the Gottesman-Kitaev-Preskill (GKP) code \cite{gottesman01}. Since the gate noise of our computation scheme is additive quadrature noise, GKP-encoding where a qubit is encoded in the mode quadratures as Dirac combs is most suitable. The GKP code is also suitable for correcting excitation loss errors since excitation loss can be converted via quantum-limited amplification into additive quadrature noise \cite{albert18,noh2019,ivan11,garcia12}. Furthermore, as the gate noise (with variance given in Eq.~\eqref{eq:gatenoise}) is added symmetrically in phase space, we consider GKP qubits encoded on square grids in phase space with a $2\sqrt{\pi}\times2\sqrt{\pi}$ unit cell. For such encoded qubits, a universal Clifford gate set is realized by the Gaussian gates $\lbrace \hat{R}(\pi/2),\hat{P}(1),\hat{C}_Z(1)\rbrace$ together with $\sqrt{\pi}$ displacements in phase space. For a comprehensive review of the GKP code, see \cite{gottesman01,tzitrin20,terhal20}.

Information encoded in GKP qubit states, $\ket{\psi_\text{in}}_\text{GKP}$, is launched into the computation scheme at the computational level as shown in Fig.~\ref{fig:setup}(a). These states are not ideal as they are subjected to the finite energy constraints (similar to the squeezed states). This means that the uncertainties of the individual spikes of the quadrature comb of the GKP state are not zero but have a finite value. Mathematically, the delta functions of the Dirac comb in the GKP state quadrature wave function are replaced by finitely squeezed Gaussian functions, each with a variance of $\sigma_\text{GKP}^2$ such that the $\hat{q}$-quadrature wave functions of the approximate GKP Pauli-Z eigenstates, $\ket{\bar{j}_\text{GKP}}$ (where $j=0,1$), are
\begin{equation}\label{eq3:wavefunction}
    \psi_j(q)\propto E(q)\sum_{n=-\infty}^\infty \!\exp\left[\frac{-(q-(2n+j)\sqrt{\pi})^2}{2\sigma_\text{GKP}^2}\right],\; j=0,1.
\end{equation}
Here $E(q)$ is an overall envelope that can be chosen to satisfy the Fourier relations between the orthogonal $\q$ and $\p$ quadratures with equally squeezed spikes---different finitely squeezed approximations of GKP states exist with different envelopes \cite{matsuura20}. In the following, for the sake of simplifying the simulation of the fault-tolerance squeezing threshold in section \ref{sec:surface}, we ignore the overall envelope, i.e. we set $E(q)=1$. Doing so corresponds to a noisier GKP state that is an incoherent mixture of ideal GKP states. This is therefore a conservative assumption that does not lead to false positive results in the noise model~\cite{noh20}. Moreover, for the small $\sigma_\text{GKP}$ values considered in this work, the envelopes are correspondingly broad, and we expect that ignoring these will have little effect on the simulated error thresholds presented here. We further assume that the squeezing of the GKP spikes in both $\hat{q}$ and $\hat{p}$ quadratures is the same as that of the $\sqv$ states in the resource preparation,
\begin{equation}\label{eq:GKPnoise}
	\sigma_\text{GKP}^2=\sigma^2=e^{-2r}/2\;.
\end{equation}

The Gaussian noise accompanying gate implementation results in the variance of the GKP spikes increasing by $\sigma_\text{gate}^2$ in both quadratures for every single gate. To prevent this, GKP quadrature correction is performed, preferably after every  gate. Traditionally, this is done by coupling each quadrature to ancillary GKP states, which are then measured, and the result is fed forward to displacements of the computational qubit (or compensated for in following measurement outcomes) \cite{gottesman01}. However, this on-demand coupling of desired modes with encoded GKP qubits to ancillary GKP states requires either active squeezing, which is experimentally hard to realize, or projective measurements, which add noise. Instead, we use the new approach by Walshe \textit{et al.} \cite{walshe20} where GKP quadrature correction is realized by qubit teleportation using ancillary GKP \emph{qunaught} states and is directly compatible with our computation scheme. The GKP qunaught state, $\qn$, is the 1-level version of the generalized GKP qu\textit{d}it state with a $\sqrt{2\pi}$ spacing between the spikes in the quadrature wave functions \cite{duivenvoorden17,walshe20}:
\[
    \psi_\varnothing(q)\propto\sum_{n=-\infty}^\infty \exp\left[\frac{-(q-n\sqrt{2\pi})^2}{2\sigma_\text{GKP}^2}\right]\;,
\]
where the overall envelope is ignored as well, and we assume the spike-squeezing in both quadratures to equal that of the GKP qubit states with variance $\sigma_\text{GKP}^2$. As such, $\qn$ holds no information, but interfering two $\qn$ states on a beam-splitter results in a two-mode GKP-qubit Bell state---for more information, see \cite{walshe20}. This state can then be used for GKP-qubit teleportation with support only on the GKP grid in phase-space so that a noisy GKP qubit is projected into a purified GKP qubit by the teleportation.

\begin{figure}
	\includegraphics[width=0.85\linewidth]{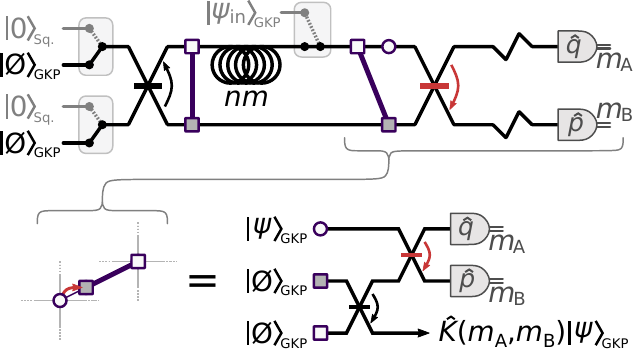}
	\caption{\label{fig:GKP} Implementation of GKP quadrature correction by qubit teleportation. Qunaught states, $\qn$, are injected in the resource preparation gadget, thereby preparing a two-mode GKP-qubit Bell state shown as two connected rectangles before the $nm$-delay. In the computational level after the $nm$-delay, one part of the Bell state overlaps in time with the GKP-qubit state to be corrected, $\ket{\psi}_\text{GKP}$, shown as a circle. Setting the TDMD to perform a Bell measurement (the two VBSs are open and left out in the figure) $\ket{\psi}_\text{GKP}$ is teleported through the Bell state and projected into a purified GKP qubit state by the Kraus operator in Eq.~\eqref{eq:Kraus}. Below is the corresponding graph as it will appear in the 3D time lattice of Fig.~\ref{fig:setup}(c) as well as the corresponding circuit diagram.}
\end{figure}

The implementation of the GKP quadrature correction in \cite{walshe20} is shown in Fig.~\ref{fig:GKP}. In the resource preparation gadget, we switch from $\sqv$ to $\qn$ states. After interference on the first beam-splitter, a GKP Bell state is prepared at the computational level instead of a two-mode CV cluster state. For teleportation of a noisy GKP qubit through the GKP Bell state, a Bell-measurement of the noisy GKP qubit and one mode of the Bell state should be carried out by the TDMD. This is done by leaving the two VBSs open and measuring in the $\q$ and $\p$ basis in spatial modes $A$ and $B$, respectively. The corresponding graph in a small section of the 3D time lattice is shown in Fig.~\ref{fig:GKP} together with the corresponding circuit. The resulting Kraus operator,
\begin{equation}\label{eq:Kraus}
	\hat{K}(m_A,m_B)=\mathcal{N}\,\hat{\bar{\Pi}}_\text{GKP}\hat{X}(-m_A\sqrt{2})\hat{Z}(-m_B\sqrt{2})\;,
\end{equation}
projects the noisy input state into a purified GKP qubit state. Here, $\hat{X}(-m_A\sqrt{2})=e^{im_A\sqrt{2}\p}$ and $\hat{Z}(-m_B\sqrt{2})=e^{-im_B\sqrt{2}\q}$ are displacements in the $\q$ and $\p$ quadratures, respectively, depending on the measurement outcomes $m_A$ and $m_B$, $\mathcal{N}$ is a normalization factor, also depending on the measurement outcomes, and
\[
	\hat{\bar{\Pi}}_\text{GKP}=\ket{\bar{0}_\text{GKP}}\bra{\bar{0}_\text{GKP}}+\ket{\bar{1}_\text{GKP}}\bra{\bar{1}_\text{GKP}}
\]
is a noisy GKP projector (here $\ket{\bar{j}_\text{GKP}}$ are the approximate GKP Pauli-Z eigenstates with the quadrature wave function in Eq.~\eqref{eq3:wavefunction}, assuming $E(q)=1$ and equal squeezing of spikes in $\hat{q}$ and $\hat{p}$ quadratures).

The output values, $m_A$ and $m_B$, are integer multiples of $\sqrt{\pi/2}$ plus some noise associated with the finite squeezing of the GKP qubit and qunaught states. As such, the $\hat{X}(-m_A\sqrt{2})$ and $\hat{Z}(-m_B\sqrt{2})$ displacements in Eq.~\eqref{eq:Kraus} mainly corresponds to Pauli-X and Pauli-Z operations on the encoded qubit, and is a natural result of the teleportation similar to regular qubit teleportation. These displacements may be compensated for by unitarily displacing the teleported state back in $\q(\p)$ by $m_{A(B)}\sqrt{2}$ rounded to the nearest integer of $\sqrt{\pi}$, or simply by shifting the final measurement outcomes. However, due to the inevitable noise in $m_A$ and $m_B$, occasionally $m_{A(B)}\sqrt{2}$ will be rounded to the wrong integer of $\sqrt{\pi}$ which then results in a faulty displacement operation. This induces a qubit error. The probability for this error to occur is \cite{noh20}
\begin{equation}\label{eq:p}
	p_\sigma(z)=\frac{\sum_{n\in\mathbb{Z}}\exp\left[-(z-(2n+1)\sqrt{\pi})^2/(2\sigma^2)\right]}{\sum_{n\in\mathbb{Z}}\exp\left[-(z-n\sqrt{\pi})^2/(2\sigma^2)\right]}
\end{equation}
where the residual analogue information, $z=\mathcal{R}(m_{A(B)}\sqrt{2})$, when rounding is given by
\begin{equation}\label{eq:R}
	\mathcal{R}(m_{A(B)}\sqrt{2})=m_{A(B)}\sqrt{2}-\sqrt{\pi}\left\lfloor\frac{m_{A(B)}\sqrt{2}}{\sqrt{\pi}}+\frac{1}{2}\right\rfloor\;,
\end{equation}
In Eq.~\eqref{eq:p}, $\sigma^2=\sigma_\text{in}^2+\sigma_\text{GKP}^2$ is the variance of $z$ with $\sigma_\text{in}^2$ being the spike variance of the GKP qubit before teleportation. For example, if the GKP qubit to be corrected has gone through one gate, then $\sigma_\text{in}^2=\sigma_\text{GKP}^2+\sigma_\text{gate}^2$ where $\sigma_\text{GKP}^2$ was the GKP qubit spike variance before the gate and  $\sigma_\text{gate}^2$ is the gate noise variance in Eq.~\eqref{eq:gatenoise}. In \cite{fukui17} it was proposed to use the analogue information from the GKP quadrature correction to improve the concatenated qubit error correction. Here, similar to \cite{noh20}, we use this analogue information through the probability in Eq.~\eqref{eq:p} to improve the second layer of error correction, the surface code, which is the subject of the next section.

\section{Surface code}\label{sec:surface}
In section \ref{sec:GKP}, we showed how to project the continuous variable noise from the finite squeezing in section \ref{sec:setup} into qubit Pauli errors by GKP quadrature correction. However, in order to perform fault-tolerant quantum computation, such Pauli errors must then be corrected using an additional quantum error correcting code operating at the qubit level. Given the nearest neighbour interactions of the computation scheme in section \ref{sec:setup}, topological qubit error correction is a natural choice to correct the Pauli errors. With information encoded on a surface of the computation scheme's 3D time lattice, and gates implemented in the third dimension, we consider the surface code \cite{bravyi98,dennis02,fowler12}. Specifically, to compute logical $X$ or $Z$ error rates, we implement the simulation methods of \cite{noh20} applied to the rotated surface code \cite{bombin07,tomita14}. Such simulation methods are adapted to the computation scheme as described in appendix \ref{app:results} and \ref{app:decoding}. We note that the rotated surface code may not be the most resource-efficient code for our computation scheme since it is rotated $45^\circ$ with respect to the 3D time lattice, and thereby, computation modes located in the corner of the 3D time lattice may not be utilized---the rotated surface code was chosen in order to easily adapt the simulation method of \cite{noh20}. Below, in section \ref{sec3:code}, we first describe the implementation of the surface code, and then consider it combined with GKP quadrature correction. In section \ref{sec3:results}, we then present simulation results of logical error rates and provide a squeezing threshold.

\begin{figure*}
	\includegraphics[width=1\linewidth]{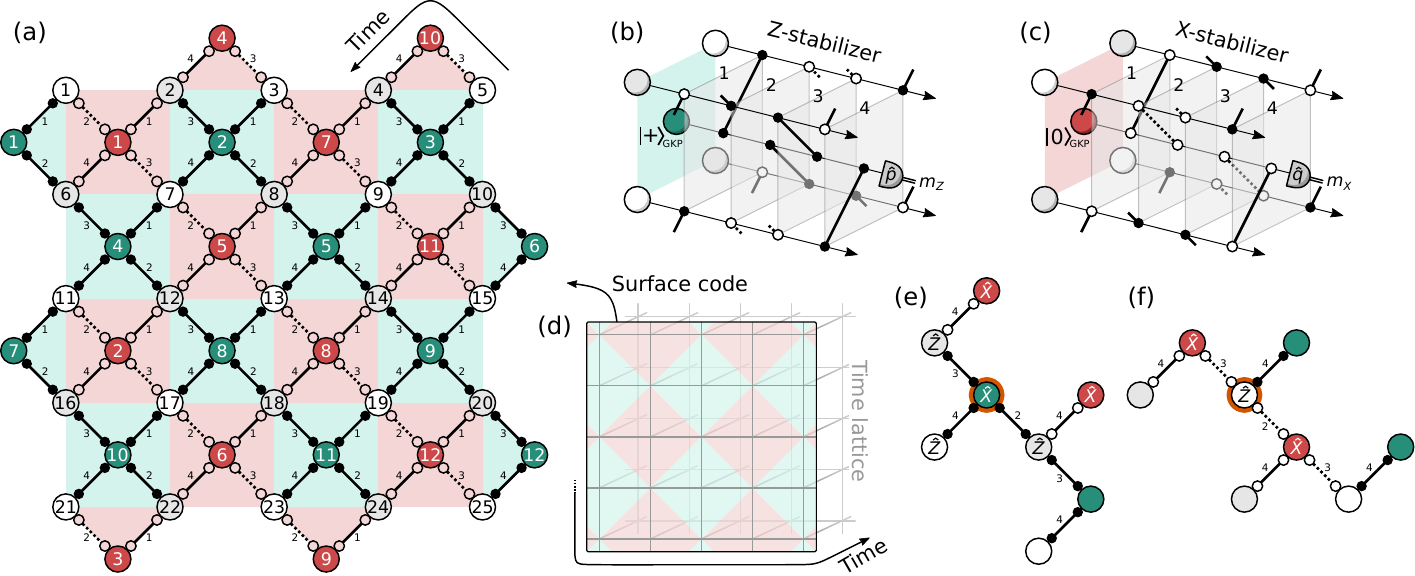}
	\caption{\label{fig:surface}(a) Illustration of a logical qubit for the $d=5$ rotated surface code. White and gray circles represent odd and even data qubits, respectively. Green and red circles represent $Z$- and $X$-measure qubits, respectively. The two-mode gate operations are listed in Eq.~\eqref{eq:surface_code_gates} and the labels 1 to 4 indicate the time steps in which those gates are implemented. (b,c) Illustration of one round of syndrome measurements including the initialization of the encoded GKP ancilla qubits, 4 time steps for the coupling between the data and ancilla qubits by 4 measurement-induced two-mode gates, and a measurement using the TDMD. For the surface-GKP code, GKP quadrature corrections are performed on data qubits at the beginning of a surface code syndrome measurement cycle. For the surface-4-GKP code, GKP quadrature corrections are performed on all data and ancilla qubits after each gate. (d) Orientation of the surface code in the 3D time lattice of the computation scheme in Fig.~\ref{fig:setup}. Here, the surface code is encoded in two dimensions of the time lattice with vertices corresponding to encoded GKP qubits, while gates are encoded in each step along the third dimension of the time lattice. (e,f) Example of qubit errors induced by GKP quadrature correction in the surface-4-GKP code. In (e), an $\hat{X}$ Pauli error occurs on a measure-$Z$ ancilla after the first two-mode gate of a surface code syndrome measurement round. The error then propagates to neighbouring data and measure-$X$ ancillas via the subsequent two-mode gates used to measure the surface code stabilizers. In (f), a $\hat{Z}$ Pauli error occurs on a data qubit after the first two-mode gate and propagates to two measure-$X$ ancillas. See appendix \ref{app:decoding} for an examination of all possible Pauli errors, how they propagate, and the corresponding edge in the matching graphs for decoding.}
\end{figure*}

\subsection{Implementation of the rotated surface code}\label{sec3:code}
A logical qubit is shown in Fig.~\ref{fig:surface}(a) for a distance $d=5$ rotated surface code. Information is encoded in $d^2$ data qubits (white and gray circles). The stabilizers of the code are measured using $(d^2-1)/2$ ancilla qubits prepared in $\ket{+}_{\text{GKP}}$ (green circles) and  $(d^2-1)/2$ ancilla qubits prepared in $\ket{0}_{\text{GKP}}$ (red circles). In what follows, we refer to green and red ancillas as measure-$Z$ and measure-$X$ ancillas. 

One round of $Z$ and $X$-type stabilizer measurements is shown in Fig.~\ref{fig:surface}(b,c). Each stabilizer measurement consists of four two-qubit gates and is thus implemented in four time steps along the third dimension of the 3D time lattice in which the surface code is implemented as shown in Fig.~\ref{fig:surface}(d). Using the optical input-switch in spatial mode $A$ of the setup described in section \ref{sec:setup}, the ancilla qubits, initialized beforehand in the $\ket{0}_\text{GKP}$ and $\ket{+}_\text{GKP}\propto\ket{0}_\text{GKP}+\ket{1}_\text{GKP}$ states, are switched into the computational level in the temporal modes corresponding to ancillary modes of the surface code. The measure-$Z$ and -$X$ ancillas are then coupled to neighbouring data qubits using $\hat{C}_Z(1)=e^{i\q\otimes\q}$ and $\hat{C}_X(\pm1)=e^{\mp i\p\otimes\p}$ gates, before being measured in the $\p$ and $\q$ basis, respectively. To measure such ancillas using the TDMD, the VBSs are left open while the same basis is chosen in spatial modes $A$ and $B$ in which case the measurements commute with the beam-splitter of the TDMD. Note that the state initialization and measurement basis for the ancillas are opposite of what is traditionally used in the surface code since they are coupled to data qubits via $\hat{C}_Z$ and $\hat{C}_X$ gates instead of sum-gates, $\hat{C}_\text{NOT}=e^{-i\q\otimes\p}$. The reason for not using sum-gates is that such gates cannot be implemented in the MBQC scheme considered in this work in a single set of projective measurements. As such, using sum-gates would lead to larger gate error rates compared to the error rates of the $\hat{C}_Z$ and $\hat{C}_X$ gates.

While the measure-$Z$ ancillas are coupled to data qubits with a constant coupling rate through $\hat{C}_Z(1)$, the measure-$X$ ancillas are coupled to data qubits with $\hat{C}_X(1)$ in step 1 and 4 and $\hat{C}_X(-1)$ in step 2 and 3. This is to prevent the propagation of finite squeezing noise among measure-qubits \cite{noh20} (though this does not matter in the case of GKP quadrature correction during the stabilizer measurements as discussed later). Furthermore, since the $\hat{C}_Z(1)$ and $\hat{C}_X(\pm1)$ gates cannot be implemented in a single computation step without Fourier by-products as described in section \ref{sec:setup}, the surface code is implemented with the two-mode gates listed in table \ref{tab:two_mode_gates}, and so, for the different two-mode gates in Fig.~\ref{fig:surface} we use
\begin{equation}\label{eq:surface_code_gates}\begin{aligned}
	\begin{matrix}
		\includegraphics[width=0.025\linewidth]{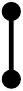}
	\end{matrix}		
	&\;\;\;=\;\;\begin{cases}
		(\hat{F}^\dagger\otimes\hat{F})\hat{C}_Z(1)\quad\;,\;\; \text{step 1 \& 3}\\
		\hat{C}_Z(1)(\hat{F}\otimes\hat{F}^\dagger)\quad\;,\;\; \text{step 2 \& 4}
	\end{cases}\;,
	\\
	\begin{matrix}
		\includegraphics[width=0.025\linewidth]{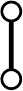}
	\end{matrix}
	&\;\;\;=\;\;\begin{cases}
		(\hat{F}\otimes\hat{F}^\dagger)\hat{C}_X(1)\quad\;,\;\; \text{step 1}\\
		\hat{C}_X(1)(\hat{F}^\dagger\otimes\hat{F})\quad\;,\;\; \text{step 4}
	\end{cases}\;,
	\\
	\begin{matrix}
		\includegraphics[width=0.025\linewidth]{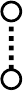}
	\end{matrix}		
	&\;\;\;=\;\;\begin{cases}
		(\hat{F}\otimes\hat{F}^\dagger)\hat{C}_X(-1)\;\;,\;\;\text{step 3}\\
		\hat{C}_X(-1)(\hat{F}^\dagger\otimes\hat{F})\;\;,\;\;\text{step 2}
	\end{cases}\;,
\end{aligned}\end{equation}
where the first term in the tensor products is the earlier temporal mode in the computational level. In this way, the Fourier by-products of step 1(3) and 2(4) cancel as $\hat{F}\hat{F}^\dagger=\hat{F}^\dagger\hat{F}=\hat{I}$ on measure-$Z$ and odd data qubits, and becomes $\hat{F}\hat{F}=\hat{F}^\dagger\hat{F}^\dagger=-\hat{I}$ on measure-$X$ and even data qubits. Hence such terms have no influence on the encoded information and do not propagate errors. For CV noise, $-\hat{I}$ on even data qubits cancels with $-\hat{I}$ on measure-$X$ qubits when phase-space displacements propagate in-between measure qubits.

We proceed to combine the surface code with GKP quadrature correction, the so-called surface-GKP code. Commonly, in the surface-GKP code, each round of syndrome measurements consists of correction of the GKP data qubits followed by measurements of the surface code stabilizers. In this way, qubit errors induced in the GKP quadrature correction is corrected by the surface code \cite{fukui18,vuillot19,noh20}. However, in the usual surface-GKP code, gate noise accumulates during all four gates of the stabilizer measurements in Fig.~\ref{fig:surface}(b,c). We propose to modify the scheme to perform GKP quadrature correction of each mode after every implemented gate. In other words, for each $Z$- and $X$-stabilizer measurement, GKP quadrature correction is performed four times, and we refer to this as the surface-4-GKP code. Unfortunately, when doing so, qubit errors are induced during the surface code stabilizer measurements with a large impact on the fault-tolerant error threshold \cite{fowler12}. Two examples of induced qubits errors, and how they propagate during the stabilizer measurements, are shown in Fig.~\ref{fig:surface}(e,f). Here, a qubit $\hat{X}$ error on a measure-$Z$ qubit, induced in the GKP quadrature correction after the first gate of the stabilizer measurement, propagates to three data qubits as $\hat{Z}$ errors through the $\hat{C}_Z$ gates (while an initial $\hat{Z}$ error will not propagate through $\hat{C}_Z$). From there it further propagates to two measure-$X$ qubits through $\hat{C}_X$ gates. Similarly, a $\hat{Z}$ error on a data qubit after the first two-mode gate propagates as $\hat{X}$ errors to measure-$X$ qubits through $\hat{C}_X$. These errors may lead to faulty syndrome measurements, and can therefore lead to wrong error recovery inducing logic errors, but even then, we will find a significant improvement of the surface-4-GKP code over the surface-GKP code. All possible Pauli errors induced by GKP quadrature correction, and their effect on the stabilizer measurements, are described in appendix C. Note, in the case here with GKP quadrature correction after every gate, having $-1$ coupling rate in $\hat{C}_X(-1)$ of step 2 and 3 is unnecessary, as all CV noise is immediately corrected. However, since a $-1$ coupling rate requires no extra resources and is solely controlled by the basis settings in table \ref{tab:two_mode_gates}, we keep it like this to compare with the surface-GKP code.

Finally, the surface code $Z$- and $X$-stabilizer measurement outcomes from $d$ rounds of syndrome measurements are recorded in the vertices of 3D $Z$ and $X$ matching graphs with edges corresponding to possible Pauli errors as described in appendix \ref{app:decoding}. For simplicity, here we consider only edges corresponding to single uncorrelated Pauli errors, and ignore possible, but less likely, two-qubit Pauli errors correlated by two-mode gates. Minimum-weight perfect matching (MWPM) \cite{edmonds65a,edmonds65b} on these matching graphs is then used as the decoding algorithm to determine data qubit errors and the resulting error recovery. In practice, the error recovery is simply handled by using and updating a Pauli frame \cite{knill05,divincenzo07,terhal15,chamberland18}, similar to how feed-forward can be handled in MBQC by compensating for by-products in the following measurement outcomes \cite{menicucci06,gu09}. For the MWPM to find error paths of highest probability, the edges of the matching graphs are dynamically weighted using Eq.~\eqref{eq:p} with the residual analogue information from each GKP quadrature correction. In this way, we can infer the probabilities of having induced the Pauli errors represented by each edge (described in appendix \ref{app:decoding}). With each edge representing multiple Pauli errors induced in different GKP quadrature corrections, multiple Pauli error probabilities are combined in each edge weight as
\begin{equation}\label{eq3:Ptot}
	p_\text{tot}=\frac{1}{2}\left(1-\prod_i\left[1-2p_i\right]\right)\;,
\end{equation}
where $p_i$ is the probability given by Eq.~\eqref{eq:p} for one GKP quadrature correction taking values between 0 (no error) and $1/2$ (minimal error information). For the combined probabilities of multiple edges to correctly add up in an error path determined by the MWPM, the edge weights in the matching graphs are finally taken to be $\log_2(p_\text{tot})$.

\subsection{Simulation results}\label{sec3:results}
\begin{figure}
	\includegraphics[width=0.75\linewidth]{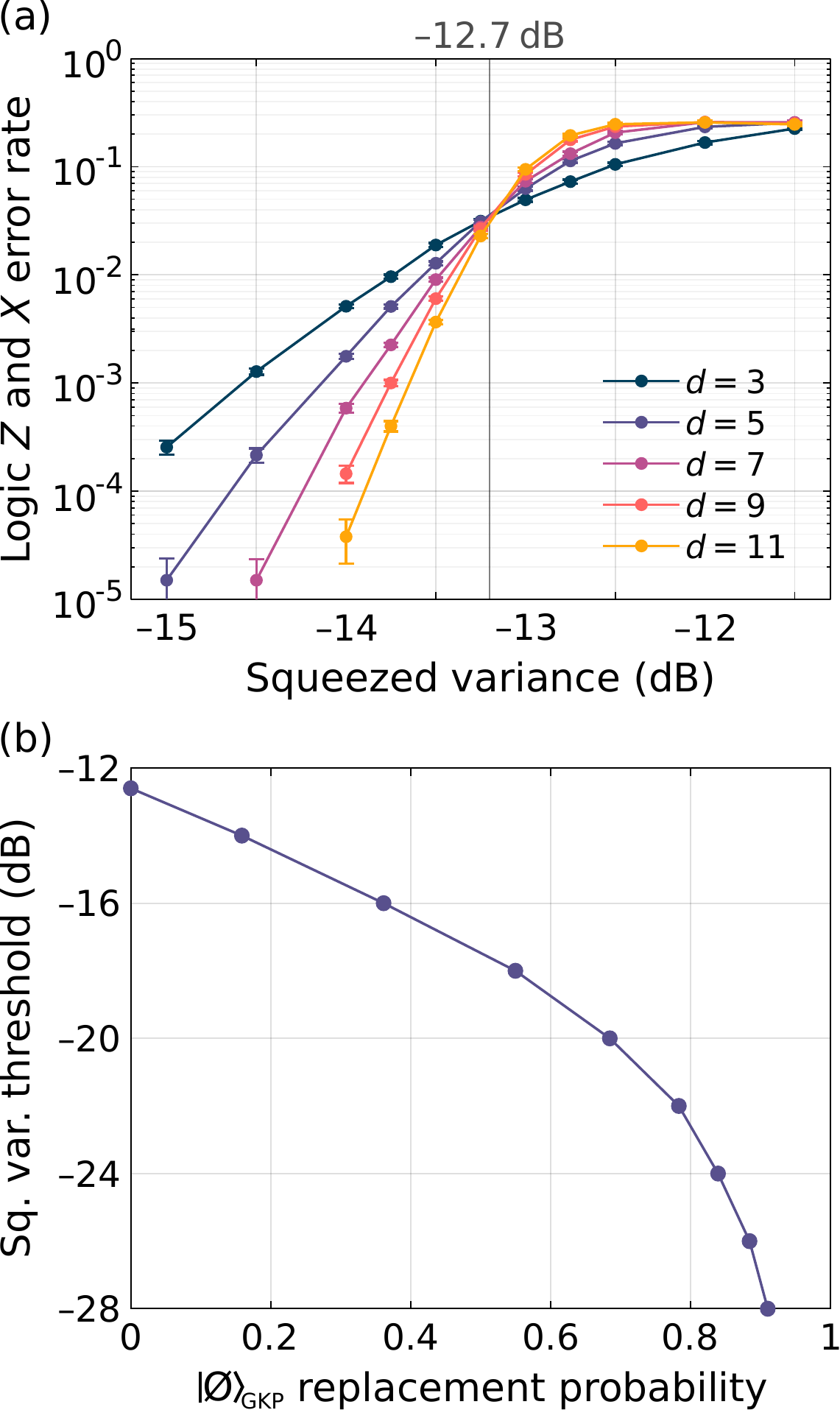}
	\caption{\label{fig:results} (a) Simulated logic $\hat{Z}$ and $\hat{X}$ error probability of the surface-4-GKP code as a function of the (identical) squeezing of the $\sqv$-states used for gate implementation, the GKP qubits encoding the surface code, and the $\qn$-states used for quadrature correction. The logic error probability is shown for different code distances $d$, and the fault-tolerant threshold where the logic error rate decreases with increasing code distance is seen to be at $\SI{12.7}{dB}$ of squeezing. Error bars of standard deviations are estimated by bootstrapping. (b) Squeezing threshold of the surface-4-GKP code as a function of the probability with which a $\qn$-state is replaced with a $\sqv$-state in the resource preparation. Here, the threshold is estimated as the crossing point of the $d=7$ and $d=9$ logic error rates. For zero replacement probability, the threshold is that of (a).}
\end{figure}

To establish a fault-tolerant error threshold, we numerically simulate the complete scheme. The GKP-encoded data and measure qubits and the qunaught states, $\qn$, are all initialized with $\sigma_\text{GKP}^2=\sigma^2=e^{-2r}/2$ variance of the wave functions' GKP spikes as described in section \ref{sec:GKP}. The ancillary squeezed vacuum states for gate implementation, $\sqv$, are as well squeezed by $\sigma^2=e^{-2r}/2$ leading to quadrature-symmetric gate noise of variance $\sigma_\text{gate}^2=2\sigma^2=e^{-2r}$ as described in section \ref{sec:setup}. Using the Monte Carlo method, logical qubit error rates are simulated as a function of squeezing using up to $\SI{100000}{}$ simulation samples with a stopping condition at the occurrence of 500 combined logic $\hat{Z}$ and $\hat{X}$ qubit error events. The resulting logical $\hat{Z}$ or $\hat{X}$ error rate (they are equal) is shown in Fig.~\ref{fig:results}(a) for different code distances $d$ as a function of squeezing level, while the logical $\hat{Y}$ error rate is smaller. The decibel scale is defined relative to the vacuum variance, $10\log_{10}[\sigma^2/(1/2)]$. The resulting squeezing threshold from where the logic error rate decrease with increasing code distance is found to be $\SI{12.7}{dB}$ of squeezing.

For comparison, in appendix \ref{app:results} we also simulate the error rates of other scenarios with the simulation results shown in Fig.~\ref{fig:all_results}. For the surface-GKP code with a single GKP quadrature correction before the surface code stabilizer measurements, the squeezing threshold increases to $\SI{17.3}{dB}$. This is significantly higher than the $\SI{12.7}{dB}$ squeezing threshold of the surface-4-GKP code due to accumulation of gate noise during the stabilizer measurements. To compare with other MBQC schemes with topological error correction where gate noise is typically not taken into account, and so only includes finite squeezing noise from GKP states, we simulate the surface-4-GKP code with $\sigma_\text{gate}^2=0$. The resulting threshold is $\SI{10.2}{dB}$ of squeezing which agrees well with the $\SI{10}{dB}$ reported in \cite{fukui18} and the $\SI{10.5}{dB}$ reported in \cite{bourassa20}. Finally, to see the impact of using the residual analogue information of the GKP quadrature correction in the weighting of the matching graphs for MWPM decoding, we simulate the surface-4-GKP code with fixed weighting based on variances of each mode at each point in the code similar to \cite{noh20}. As expected, the result is a slightly larger squeezing threshold of $\SI{13.6}{dB}$.

While the GKP code has been experimentally realized in trapped-ion and circuit QED systems \cite{Fluhmann2019,campagne20,deNeeve2020}, GKP state generation in optical platforms is yet to be demonstrated, although there are many recent proposals \cite{motes17,vasconcelos10,weigand18,eaton19,SCC19,su19,tzitrin20}. At first, GKP state generation is most likely going to be probabilistic. In Ref.~\cite{bourassa20} it is proposed to combine multiple GKP state generators with optical switches, and then switch between generators with a successful preparation of a GKP state. In this way, the success probability of the GKP state generation, $p_\varnothing$, can in principle be brought arbitrarily close to 1. Since the surface-4-GKP code requires a large supply of $\qn$ states, we consider as our final analysis the multi-GKP state generation scheme of Ref.~\cite{bourassa20} for  $\qn$ resource state preparation. If all the generators fail to prepare a $\qn$ state in a given temporal mode for GKP correction, a deterministically generated squeezed vacuum state, $\sqv$, is used instead. In this case, if $\qn$ is replaced by $\sqv$ in spatial mode $A$($B$), only the $\q$($\p$) quadrature is corrected in the GKP quadrature correction, while the other quadrature accumulates gate noise of variance $\sigma_\text{gate}^2$ during the correction \cite{walshe20}. The resulting fault-tolerance squeezing threshold is shown in Fig.~\ref{fig:results}(b) as a function of the probability $1-p_\varnothing$ of replacing $\qn$ states by $\sqv$ states. Here, a squeezing threshold is seen to exist in a large range of $1-p_\varnothing>0$, allowing for a probabilistic supply of $\qn$ states, while for increasing replacement probability, the squeezing threshold level increases as expected---above $\SI{28}{dB}$ squeezing it is hard for us to simulate the squeezing threshold due to low error rates of less than $10^{-7}$. Note, here we still assume successful encoding of the surface code. I.e., the data and measure qubits switched into the setup as $\ket{\psi_\text{in}}_\text{GKP}$ in Fig.~\ref{fig:setup}(a) are successfully prepared as GKP qubit states. With probabilistic optical GKP state generation, this may be possible using state storage of a probabilistically prepared GKP state until it is switched into the computation scheme \cite{lvovsky09,zhong17,bouillard19,hashimoto19}.

\section{Discussion}\label{sec:discussion}
The squeezing thresholds in this work are derived by assuming a particular noise model in which all resource states are finitely squeezed while all optical propagation and detection losses are set to zero. In practice, however, losses cannot be neglected. Let us denote the transmission of the setup by $\eta$. For Gaussian states, $0 < \eta<1$  leads to the formation of mixed states with reduced effective squeezing. This can be reformulated as an ideal, loss-less setup ($\eta=1$) with mixed squeezed vacuum states as input having a lower effective squeezing, and some excess anti-squeezing that does not affect the measurement-based computation \cite{walshe19}. As a result, for $\eta<1$ the $\SI{12.7}{dB}$ squeezing threshold corresponds to the effectively measured squeezing. Now, for the GKP states, besides a Gaussian convolution in the quadratures, $\eta<1$ leads to a ``shrinking'' of a GKP state in phase-space. To see this, consider the Heisenberg picture with $\eta$ modeled as a beam-splitter of $\eta$ transmission. In this case, an amount $1-\eta$ of vacuum is mixed into the state, adding noise to the quadratures, while a share $1-\eta$ of the state is lost, ``shrinking'' the state in the quadratures by $\sqrt{\eta}$. The quadrature shrinking is more detrimental to GKP spikes far from the phase-space origin, which are naturally delimited in GKP states of finite squeezing due to the overall envelope in the quadrature wave function. For GKP-states with $\SI{12.7}{dB}$ of squeezing, we assume this effect to be negligible on the qubit error probabilities for reasonably high efficiencies---we estimate $\eta\gtrsim0.95$ to be doable on optical platforms. We also note that the shrinking effect can be counteracted by linear amplification which on the other hand will further reduce the amount of squeezing \cite{albert18,noh2019,ivan11,garcia12}, effectively resetting $\eta$ to unity at the cost of lowering the effective squeezing and the purity of the GKP state. Again, the estimated threshold of $\SI{12.7}{dB}$ refers to the required squeezing after such actions have been implemented. Another detrimental effect that has not been directly accounted for is interferometric phase fluctuations. Similar to optical loss, phase fluctuations lead to mixed squeezed states of reduced squeezing and excess anti-squeezing as well as mixed GKP states with an impact that increases with the quadrature value. 

Finally, we comment on the scalability of the computation scheme. For the temporal encoding in Fig.~\ref{fig:setup}, the number of modes in which GKP qubits can be encoded for computation, i.e.\ the size of the encoding plane in the 3D time lattice, depends on the $nm$-delay in the resource preparation gadget. Increasing the delay length increases the number of encoding modes. However, doing so also increases the optical propagation loss, which puts a limit on the useful delay length. Thus, to continue scaling up, $nm$ must be increased by shortening the temporal modes, in turn increasing the demands on the squeezing and detection bandwidth. In \cite{kashiwazaki20}, squeezed light with a bandwidth of $\SI{2.5}{THz}$ was demonstrated, limited by the phase-matching condition of the non-linear down conversion process, while in \cite{takanashi20}, detection of squeezing up to $\SI{3}{THz}$ sideband frequency was demonstrated. Assuming proper squeezing, experimental control, and detection in a $\SI{2.5}{THz}$ bandwidth defining temporal modes of $\sim1/\SI{2.5}{THz}$ duration, and assuming a propagation efficiency above 0.95 ($\SI{0.23}{dB}$ attenuation) in a low-loss optical fiber with low optical attenuation of $\SI{0.15}{dB/km}$, up to  $nm\approx10^7$ computation modes may be realized in the temporally encoded computation scheme. For the spatial architecture of Fig.~\ref{fig:spatial}, scalability is similar to other schemes based on spatial encoding. It relies on the availability of resources, and is suitable for integrated photonics \cite{wang20}. Finally, temporal and spatial encoding may be combined: Consider multiple temporally encoded computational devices, each as in Fig.~\ref{fig:setup}. Using the optical switch at the setup computational level, computation modes can be switched in and out between different devices. Since the setup is optical, the devices are simply connected by optical fibers between the switches of each device without the need of quantum transducers. Furthermore, with the switch being mode selective, each mode of an encoded logical qubit in the surface code can be transferred without the need of decoding and re-encoding the logical quantum state, while measurement of the surface code stabilizers after transfer may be used for error-correcting the transfer line. This is not only suitable for combining temporal and spatial encoding for up-scaling, but is also useful in a quantum internet scheme \cite{kimble08,wehner18}, and is made possible by the optical architecture combined with temporal multiplexing on the transfer lines.

\section{Conclusion}\label{sec:conclusion}
In this work, we have proposed a simple but complete and scalable architecture for optical CV MBQC that includes quadrature noise correction and qubit error correction using topological codes. The setup consists of simple optical devices such as beam-splitters, delays, optical switches, and variable beam-splitters, where the latter two can be decomposed into beam-splitters and optical phase shifters. The scheme allows for both spatial and temporal encoding, with the temporally encoded version requiring just two squeezing sources. A universal Gaussian gate set is directly implementable, while universal qubit computation is made possible by feeding the setup with GKP states, thereby supplying the required non-Gaussianity \cite{baragiola19,yamasaki19,hastrup20}. As the computation scheme is based on gate teleportation on wires of two-mode entangled states, the setup naturally supports the new GKP quadrature correction scheme in Ref.~\cite{walshe20}, circumventing the need for on-line two-mode gates coupling to ancillary GKP states. Finally, by arranging the GKP qubits in a 2D plane of the cluster state that allows for nearest-neighbor interactions, topological codes can be realized. By encoding a variation of the surface-GKP code---the surface-4-GKP code---we show fault-tolerant computation to be possible above a certain squeezing threshold by simulating a logic qubit memory, or an identity gate, of the surface-4-GKP code. In the surface code, Clifford gates can be implemented by braiding \cite{fowler12} or lattice surgery \cite{horsman12} implemented by regulating the surface code syndrome measurements, while non-Clifford gates may be realized using magic states distilled from GKP qubits prepared in a magic state \cite{baragiola19,yamasaki19} and injected into the surface code as input states \cite{fowler12,litinski19}.

The fault-tolerant squeezing threshold is found to be $\SI{12.7}{dB}$. The estimation of this number takes into account the finite squeezing values of GKP states as well as the gate noise stemming from the finite squeezing values of the generated cluster state on which gates are implemented by projective measurements. However, this squeezing threshold leaves room for improvements: In the matching graph of the decoding algorithm, we have only considered single uncorrelated Pauli errors. As an improvement, considering the surface code to consist of two-mode gates, we may consider a matching graph taking two-qubit error events into account \cite{FowlerEdgeWeights,ChamberlandPRX,CKYZ20,AWSpaper2020}. Furthermore, when weighting the matching graph edges with analogue information from GKP quadrature correction, we simply consider uncorrelated noise in the individual GKP corrections. However, since each mode is subjected to a two-mode gate prior to GKP correction leading to correlated noise on neighbouring GKP qubits, we may improve the estimation of Pauli error probabilities used in the matching graph weighting by jointly considering the analogue information from neighbouring GKP corrections as recently proposed in Ref.~\cite{noh21}. Note, that such modifications are solely implemented at the software level of the error correction decoder and thus requires no modifications to the setup. Another improvement may be found in the gate implementation: Due to the similarity of the GKP quadrature correction and gate implementation, it might be possible to combine the two transformations in one step, that is, implementing a gate while correcting the quadratures. Although the quadrature correction is only considered on single wires~\cite{walshe20}, it might be possible to generalize it to the two-wire case by which two-mode gates could be implemented during GKP quadrature correction, thereby eliminating gate noise coursed by finite squeezing. If this is possible while maintaining the GKP quadrature correction quality of Ref.~\cite{walshe20}, the resulting squeezing threshold reduces to $\SI{10.2}{dB}$ as shown in appendix \ref{app:results}.

	\begin{acknowledgments}
		The work was supported by the Danish National Research Foundation through the Center for Macroscopic Quantum States (bigQ, DNRF0142).
    \end{acknowledgments}

	\appendix
	
\section{Gates by projective measurements}\label{app:gates}
To derive the gates implemented by projective measurements in section \ref{sec:setup}, consider one computation step on two parallel wires,
\begin{equation}\label{eq:two_wires}
    \begin{matrix}
    	\includegraphics[width=0.45\linewidth]{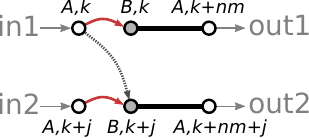}\;,
    \end{matrix}
\end{equation}
where the red arrows represent the first beam-splitter in the TDMD, and the gray arrow represents the blue or green VBS in the TDMD for $j=1$ or $j=n$, respectively. Here, a two-mode input state (separable or not) is encoded in modes $(A,k),(A,k+j)$, while the two--mode entangled states are prepared in the resource preparation gadget, and can be written as
\begin{equation*}
	\begin{matrix}
	    \includegraphics[width=0.85\linewidth]{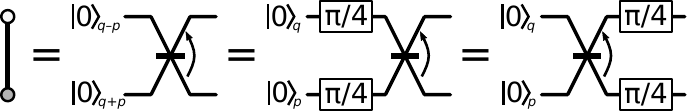}\;,
	\end{matrix}
\end{equation*}
where $\ket{0}_{q-p}$ and $\ket{0}_{q+p}$ are squeezed vacuum states, squeezed along $(\hat{q}-\hat{p})/\sqrt{2}$ and $(\hat{q}+\hat{p})/\sqrt{2}$ quadratures, respectively, and similarly, $\ket{0}_q$ and $\ket{0}_p$ are squeezed along the $\hat{q}$ and $\hat{p}$, respectively. As such, the two-mode entangled states correspond to two-mode squeezed states rotated in phase-space by $\pi/4$, turning them into approximate cluster states with $\tanh 2r$ edge weight and $i\text{sech} 2r$ self-loops, where $r$ is the squeezing parameter of the initial squeezed vacuum states \cite{menicucci11a,wu19}. Alternatively, we can consider the two-mode entangled states more generally as cluster-type states \cite{vanloock07}, here with edge weight 1, for which the implemented gate is independent on the squeezing, $r$, which then only affects the gate noise \cite{larsen20b}. The two situations are equivalent: One can change from the former to the latter by normalizing the edge weight \cite{alexander14}. Here, we will consider cluster-type states, since implementing a desired gate in practice (without considering the resulting gate noise) then requires no prior knowledge of the squeezing level.

For single-mode gates the VBSs of the TDMD are left open, and the dashed arrow in Eq.~\eqref{eq:two_wires} represents $\hat{I}_{A,k}\otimes\hat{I}_{B,k+j}$. In this case, we can ignore the second wire, and focus on a joint projective measurement of the input mode $(A,k)$ and one mode of the two-mode entangled state, $(B,k)$, resulting in gate teleportation to the output mode $(A,k+nm)$---exactly the same derivation can be made on the second wire of mode $(A,k+j)$, $(B,k+j)$, and $(B,k+nm+j)$. The corresponding circuit is
\begin{equation*}
	\includegraphics[width=0.75\linewidth]{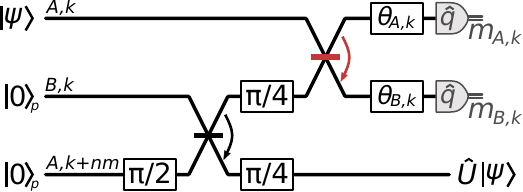}
\end{equation*}
where $\ket{0}_q$ in $(A,k+nm)$ is replaced by $\hat{R}(\pi/2)\ket{0}_p$ only to follow the traditional convention of cluster states with initial squeezing in $\hat{p}$ quadratures. Using the method of appendix A in Ref.~\cite{larsen20} with $(\theta_{A,k},\theta_{B,k})$ being the basis setting determining the implemented gate, the corresponding quadrature transformation in the Heisenberg picture can be derived to be
\begin{equation*}
	\begin{pmatrix}
		\q_{A,k+nm}'\\ \p_{A,k+nm}'
	\end{pmatrix}
	=\textbf{G}
	\begin{pmatrix}
			\q_{A,k}\\ \p_{A,k}
		\end{pmatrix}
	+\textbf{N}
	\begin{pmatrix}
		\p_{B,k}\\ \p_{A,k+nm}
	\end{pmatrix}
	+\textbf{D}
	\begin{pmatrix}
		m_{A,k}\\ m_{B,k}
	\end{pmatrix}.
\end{equation*}
Here, $\textbf{G}$ is the symplectic matrix corresponding to the desired single-mode gate operation in Eq.~\eqref{eq2:signle_mode_gate},
\begin{equation*}
	\textbf{N}=
	\begin{pmatrix}
		1 & 1\\ 1 & -1
	\end{pmatrix}
\end{equation*}
is a gate noise matrix, and 
\begin{equation*}
	\textbf{D}=\frac{\sqrt{2}}{\sin(2\theta_-)}
	\begin{pmatrix}
		-\cos\theta_{B,k} & -\cos\theta_{A,k}\\ \sin\theta_{B,k} & \sin\theta_{A,k}
	\end{pmatrix}
\end{equation*}
is a displacement matrix. Since $(\theta_{A,k},\theta_{B,k})$, $m_{A,k}$ and $m_{B,k}$ are known, $\textbf{D}(m_{A,k},m_{B,k})^T$ can be compensated for by displacing the teleported state back by $-\textbf{D}(m_{A,k},m_{B,k})^T$, or simply by taking this displacement into account in the following measurement outcomes. 
With finite squeezing in the ancillary modes such that $\text{Var}\{ \p_{B,k} \}=\text{Var}\{ \p_{A,k+nm} \}=\sigma^2=e^{-2r}/2$, the noise term $\textbf{N}(\p_{B,k},\p_{A,k+nm})^T$ leads to quadrature-symmetric gate noise in $\q_{A,k+nm}'$ and $\p_{A,k+nm}'$ of
\begin{equation*}
	\sigma_\text{gate}^2=\text{Var}\{ \p_{B,k} \}+\text{Var}\{ \p_{A,k+nm} \}=e^{-2r}\;.
\end{equation*}
In the Wigner function picture, this gate noise corresponds to convolutions in both quadratures by a Gaussian function of variance $\sigma_\text{gate}^2$, each followed by the application of a corresponding Gaussian envelope due to the Fourier relation between $\q$ and $\p$ \cite{alexander14,larsen20}.

When implementing two-mode gates by enabling the first or second VBS of the TDMD, the corresponding circuit is
\begin{equation*}
	\includegraphics[width=.95\linewidth]{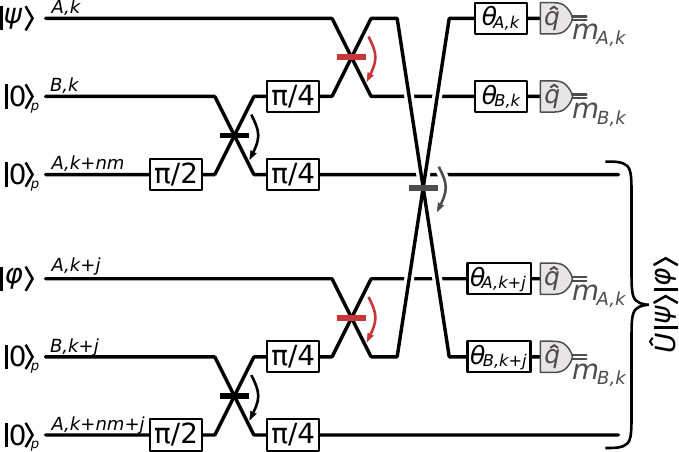}.
\end{equation*}
We do not derive a general expression for the implemented gate as a function of the basis setting $(\theta_{A,k},\theta_{B,k},\theta_{A,k+j},\theta_{B,k+j})$. Instead, we use the method described in Ref.~\cite{larsen20}: A cost function is defined based on the implemented gate, a desired target gate, and the gate noise, which is then used in a global search to find the basis setting that implements a desired gate with the minimum gate noise. The resulting basis settings for the gates required to implement the surface code are shown in Table \ref{tab:two_mode_gates}. We note that those settings only implement two different gates since $(\hat{F}^\dagger\otimes\hat{F})\hat{C}_Z(g)=\hat{C}_X(g)(\hat{F}^\dagger\otimes\hat{F})$ and $\hat{C}_Z(g)(\hat{F}\otimes\hat{F}^\dagger)=\hat{C}_X(g)(\hat{F}^\dagger\otimes\hat{F})$. The reason for considering them as four different gates is to make the implementation of the surface code more intuitive. The basis settings are not unique: other settings exist that implement the same gates with equal gate noise.

The quadrature transformation when applying the basis settings for two-mode gates is
\begin{equation*}\begin{aligned}
	&\begin{pmatrix}
		\q_{A,k+nm}'\\ \q_{A,k+nm+j}'\\ \p_{A,k+nm}'\\ \p_{A,k+nm+j}'
	\end{pmatrix}
	=\textbf{G}
	\begin{pmatrix}
			\q_{A,k}\\ \q_{A,k+j}\\ \p_{A,k}\\ \p_{A,k+j}
	\end{pmatrix}\\
	&\quad\quad\quad\quad+\textbf{N}
	\begin{pmatrix}
		\p_{B,k}\\ \p_{A,k+nm}\\ \p_{B,k+j}\\ \p_{A,k+nm+j}
	\end{pmatrix}
	+\textbf{D}
	\begin{pmatrix}
		m_{A,k}\\ m_{B,k}\\ m_{A,k+j}\\ m_{B,k+j}
	\end{pmatrix}.
\end{aligned}\end{equation*}
Again, $\textbf{G}$ is the symplectic matrix corresponding the implemented two-mode gate. $\textbf{D}(m_{A,k},m_{B,k},m_{A,k+j},m_{B,k+j})^T$ is a displacement in phase-space with
\begin{equation*}
	\textbf{D}=\begin{pmatrix}
		-\sqrt{5}/2 & 0 & 1/\sqrt{2} & \sqrt{5}/2\\
		-\sqrt{5}/2 & -1/\sqrt{2} & 0 & -\sqrt{5}/2\\
		0 & -\sqrt{2} & 0 & 0\\
		0 & 0 & \sqrt{2} & 0
	\end{pmatrix}
\end{equation*}
for $(\hat{F}^\dagger\otimes\hat{F})\hat{C}_Z(g)$ and $\hat{C}_X(g)(\hat{F}^\dagger\otimes\hat{F})$, and
\begin{equation*}
	\textbf{D}=\begin{pmatrix}
		0 & -\sqrt{2} & 0 & 0\\
		0 & 0 & -\sqrt{2} & 0\\
		\sqrt{5}/2 & 0 & -1/\sqrt{2} & \sqrt{5}/2\\
		\sqrt{5}/2 & -1/\sqrt{2} & 0 & -\sqrt{5}/2
	\end{pmatrix}
\end{equation*}
for $\hat{C}_Z(g)(\hat{F}\otimes\hat{F}^\dagger)$ and $\hat{C}_X(g)(\hat{F}^\dagger\otimes\hat{F})$, both of which can be compensated for, just like for single-mode gates. $\textbf{N}(\p_{B,k},\p_{A,k+nm},\p_{B,k+j},\p_{A,k+nm+j})^T$ represents gate noise where 
\begin{equation*}
	\textbf{N}=\begin{pmatrix}
		1 & 1 & 0 & 0\\
		0 & 0 & 1 & 1\\
		1 & -1 & 0 & 0\\
		0 & 0 & 1 & -1
	\end{pmatrix}
\end{equation*}
leads to quadrature-symmetric gate noise of variance
\begin{equation*}\begin{aligned}
	\sigma_\text{gate}^2=&\text{Var}\{ \p_{B,k} \}+\text{Var}\{ \p_{A,k+nm} \}\\
	=&\text{Var}\{ \p_{B,k+j} \}+\text{Var}\{ \p_{A,k+nm+j} \}\\
	=&e^{-2r}\;,
\end{aligned}\end{equation*}
which conveniently equals the gate noise variance of single-mode gates.

\section{Simulation}\label{app:results}
To simulate the logic qubit error rate of the surface code, we adopted and modified the simulation in Ref.~\cite{noh20} to the computation scheme of this work. The simulation method is well-described in  appendix B of Ref.~\cite{noh20} and is summarized here with focus on the modifications. In the simulation, quadrature noise is simulated as stochastic normally-distributed variables for each quadrature of each mode $i$, $\xi_q^i$ and $\xi_p^i$. For GKP-states, $\xi_q^i$ and $\xi_p^i$ are initialized with random samples  from $\mathcal{N}(0,\sigma_\text{GKP})$, where $\mathcal{N}(0,\sigma)$ is a normal distribution of zero mean and $\sigma^2$ variance. After each gate, independent random samples from $\mathcal{N}(0,\sigma_\text{gate})$ are added to $\xi_q^i$ and $\xi_p^i$ as gate noise. As for homodyne measurements, $\xi_q^i$ or $\xi_p^i$ is read out, and the logic value is determined from the closest integer multiple of $\sqrt{\pi}$. Note that, unlike in Ref.~\cite{noh20}, we do not consider measure noise or idle noise. In optical platforms, homodyne measurements are carried out with near-unity efficiency (any loss is assumed to just degrade the squeezing as discussed in section~V). Furthermore, in MBQC no modes are idle since modes not performing any tasks still have to teleport through the computation step and thereby acquire gate noise instead of idle noise. 

For the two-mode gates in the surface code, the simulation here differs from Ref.~\cite{noh20} by using $\hat{C}_Z(1)$ and $\hat{C}_X(\pm1)$ gates instead of sum-gates. For a two-mode gate between modes $i$ and $j$, the quadrature noise variables are updated as
\begin{equation*}\begin{aligned}
\hat{C}_Z(1):\quad
	&\begin{matrix*}[l]
		\xi_q^i\leftarrow\xi_q^i&+\text{ randG}(\sigma_\text{gate}^2)\\[3pt]
		\xi_q^i\leftarrow\xi_p^i+\xi_q^j&+\text{ randG}(\sigma_\text{gate}^2)\\[3pt]
		\xi_q^j\leftarrow\xi_q^j&+\text{ randG}(\sigma_\text{gate}^2)\\[3pt]
		\xi_q^j\leftarrow\xi_p^j+\xi_q^i&+\text{ randG}(\sigma_\text{gate}^2)
	\end{matrix*}\;,\\[10pt]
\hat{C}_X(\pm1):\quad
	&\begin{matrix*}[l]
		\xi_q^i\leftarrow\xi_q^i\pm\xi_p^j&+\text{ randG}(\sigma_\text{gate}^2)\\[3pt]
		\xi_q^i\leftarrow\xi_p^i&+\text{ randG}(\sigma_\text{gate}^2)\\[3pt]
		\xi_q^j\leftarrow\xi_q^j\pm\xi_p^i&+\text{ randG}(\sigma_\text{gate}^2)\\[3pt]
		\xi_q^j\leftarrow\xi_p^j&+\text{ randG}(\sigma_\text{gate}^2)
	\end{matrix*}\;,
\end{aligned}\end{equation*}
where $\text{randG}(\sigma^2)$ returns a random value from $\mathcal{N}(0,\sigma)$.

For GKP quadrature correction, instead of coupling to ancillary GKP qubits through sum gates as in Ref.~\cite{noh20}, the mode to be corrected is teleported through a two-mode GKP qubit Bell state as described in section \ref{sec:GKP}. The Bell state is prepared by interfering two GKP qunaught states, denoted $\varnothing1$ and $\varnothing2$,
\begin{equation*}
	\begin{matrix*}[l]
		\xi_q^{\varnothing1}=\text{ randG}(\sigma_\text{GKP}^2)\\[3pt]
		\xi_p^{\varnothing1}=\text{ randG}(\sigma_\text{GKP}^2)\\
	\end{matrix*}\quad,\quad
	\begin{matrix*}[l]
		\xi_q^{\varnothing2}=\text{ randG}(\sigma_\text{GKP}^2)\\[3pt]
		\xi_p^{\varnothing2}=\text{ randG}(\sigma_\text{GKP}^2)
	\end{matrix*}
\end{equation*}
on a beam-splitter,
\begin{equation*}
	\begin{matrix*}[l]
		\xi_q^{\varnothing1}\leftarrow(\xi_q^{\varnothing1}-\xi_q^{\varnothing2})/\sqrt{2}\\[3pt]
		\xi_p^{\varnothing1}\leftarrow(\xi_p^{\varnothing1}-\xi_p^{\varnothing2})/\sqrt{2}\\[3pt]
		\xi_q^{\varnothing2}\leftarrow(\xi_q^{\varnothing1}+\xi_q^{\varnothing2})/\sqrt{2}\\[3pt]
		\xi_p^{\varnothing2}\leftarrow(\xi_p^{\varnothing1}+\xi_p^{\varnothing2})/\sqrt{2}\;.
	\end{matrix*}
\end{equation*}
To teleport, the mode to be corrected, $i$, and $\varnothing1$ are interfered on a beam-splitter and measured in $\q$ and $\p$, respectively, with outcomes
\begin{equation*}
		m_A=(\xi_q^i-\xi_q^{\varnothing1})/\sqrt{2}\quad,\quad m_B=(\xi_p^i+\xi_p^{\varnothing1})/\sqrt{2}\;.
\end{equation*}
Finally, to compensate for the Pauli by-products of the qubit teleportation (displacements by $\sqrt{\pi}$), $m_A\sqrt{2}$ and $m_B\sqrt{2}$ are rounded to the nearest integer multiple of $\sqrt{\pi}$,
\begin{equation}\label{eq:P}
	\mathcal{P}(m_{A(B)}\sqrt{2})=\sqrt{\pi}\left\lfloor\frac{m_{A(B)}\sqrt{2}}{\sqrt{\pi}}+\frac{1}{2}\right\rfloor\;,
\end{equation}
which is then used to displace the teleportation output mode, $\varnothing2$, back,
\begin{equation*}
	\begin{matrix*}[l]
		\xi_q^{\varnothing2}\leftarrow\xi_q^{\varnothing2}+\mathcal{P}(m_A\sqrt{2})\\[3pt]
		\xi_p^{\varnothing2}\leftarrow\xi_p^{\varnothing2}+\mathcal{P}(m_B\sqrt{2})\;.
	\end{matrix*}
\end{equation*}
For the sake of simulation, we pass the corrected output mode to the input mode, $\xi_q^i\leftarrow\xi_q^{\varnothing2}$ and $\xi_p^i\leftarrow\xi_p^{\varnothing2}$, such that mode $i$ can be reused in the following simulation. The probability of having induced a qubit error by rounding to a wrong integer of $\sqrt{\pi}$ in Eq.~\eqref{eq:P} due to input noise in $\xi_q^i$ and $\xi_p^i$, together with initialization noise of $\xi_q^{\varnothing1}$, $\xi_p^{\varnothing1}$, $\xi_q^{\varnothing2}$, and $\xi_p^{\varnothing2}$, is inferred using the residual analogue information, $\mathcal{R}(m_{A(B)}\sqrt{2})=m_{A(B)}\sqrt{2}-\mathcal{P}(m_{A(B)}\sqrt{2})$, in Eq.~\eqref{eq:R} through the probability in Eq.~\eqref{eq:p}. Finally, it is used for weighting the matching graphs of stabilizer measurement outcomes for the MWPM decoding as described in appendix \ref{app:decoding}. Here, $\sigma_\text{in}^2$ in $\sigma^2=\sigma_\text{in}^2+\sigma_\text{GKP}^2$ of Eq.~\eqref{eq:p} is the quadrature variance of the input mode, and is carefully kept track of in the simulation based on previous gates and corrections.

In one simulation, $d+1$ rounds of surface code stabilizer measurements are carried out. Data GKP qubits are initialized in round $1$ with $\sigma_\text{GKP}^2$ variance. To stabilize the data qubits, measure GKP qubits and qunaught states are initialized in round $1$ to $d$ with $\sigma_\text{GKP}^2$ variance, followed by noisy gates and measurements to build up the matching graphs. In the last round, $d+1$, measure GKP qubits and qunaught states are initialized with zero variance to carry out ideal syndrome measurements for determining logic qubit errors induced in round $1$ to $d$. To build up statistics, for each squeezing level and code distance, $d$, this process is repeated $\SI{100000}{}$ times, or until a total of 500 logic $\hat{X}$ and $\hat{Z}$ errors are detected.

\begin{figure*}
	\includegraphics[width=0.8\textwidth]{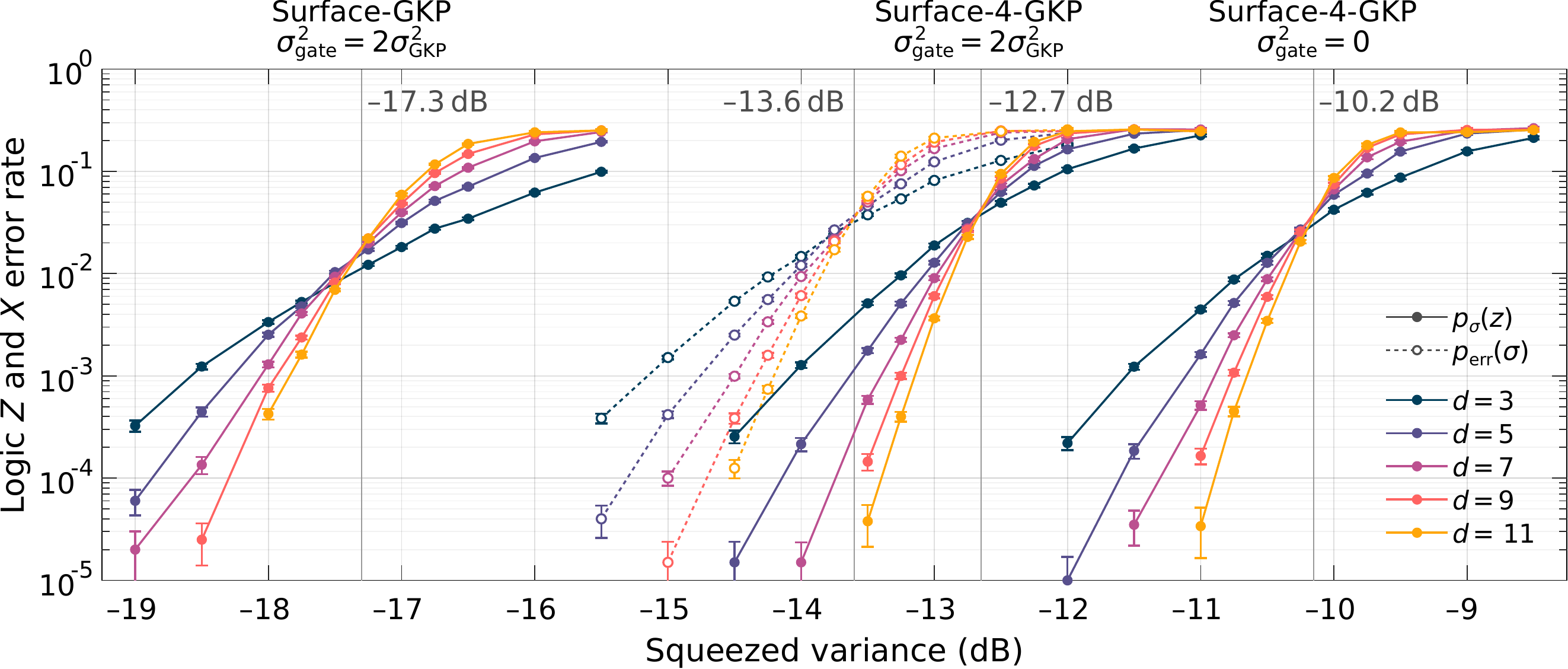}
	\caption{\label{fig:all_results} Simulation results of all four simulated cases. Here, the case of the surface-4-GKP code using $p_\sigma(z)$ is the results shown in the main text Fig.~\ref{fig:results}. Error bars of standard deviation are estimated by bootstrapping.}
\end{figure*}	

We have simulated four different cases, shown in Fig.~\ref{fig:all_results}. In three cases, GKP states (qubits and qunaught states) and squeezed vacuum states are initialized with equal variance, $\sigma_\text{GKP}^2=\sigma^2=e^{-2r}/2$, which from  Eq.~\eqref{eq:gatenoise} and \eqref{eq:GKPnoise} leads to $\sigma_\text{gate}^2=2\sigma_\text{GKP}^2$. In this way, the surface-GKP code with GKP quadrature correction before the surface code stabilizer measurements, and the surface-4-GKP code with four GKP quadrature corrections during the stabilizer measurements, was simulated. To see the impact of using the analogue information from the GKP correction in the weighting of the matching graphs, the surface-4-GKP code was simulated using 
\begin{equation}
	p_\text{err}(\sigma)=\sum_{n\in\mathbb{Z}}\frac{1}{\sqrt{2\pi\sigma^2}}\int_{(2n+\frac{1}{2})\sqrt{\pi}}^{(2n+\frac{3}{2})\sqrt{\pi}}d\xi\,e^{-\xi^2/(2\sigma^2)}
\end{equation}
instead of Eq.~\eqref{eq:p} \cite{noh20}. By integrating the wave function marginal distribution in the odd GKP bins, $p_\text{err}(\sigma)$ infers the qubit error probability only based on variances without taking the projective measurement outcome into account. Finally, to compare with other MBQC schemes supporting topological error correction, but only taking noise from GKP-states into account, the surface-4-GKP code is simulated using $\sigma_\text{gate}^2=0$.

\section{Decoding graphs}\label{app:decoding}
\begin{figure}
	\centering
	\includegraphics[width=0.75\linewidth]{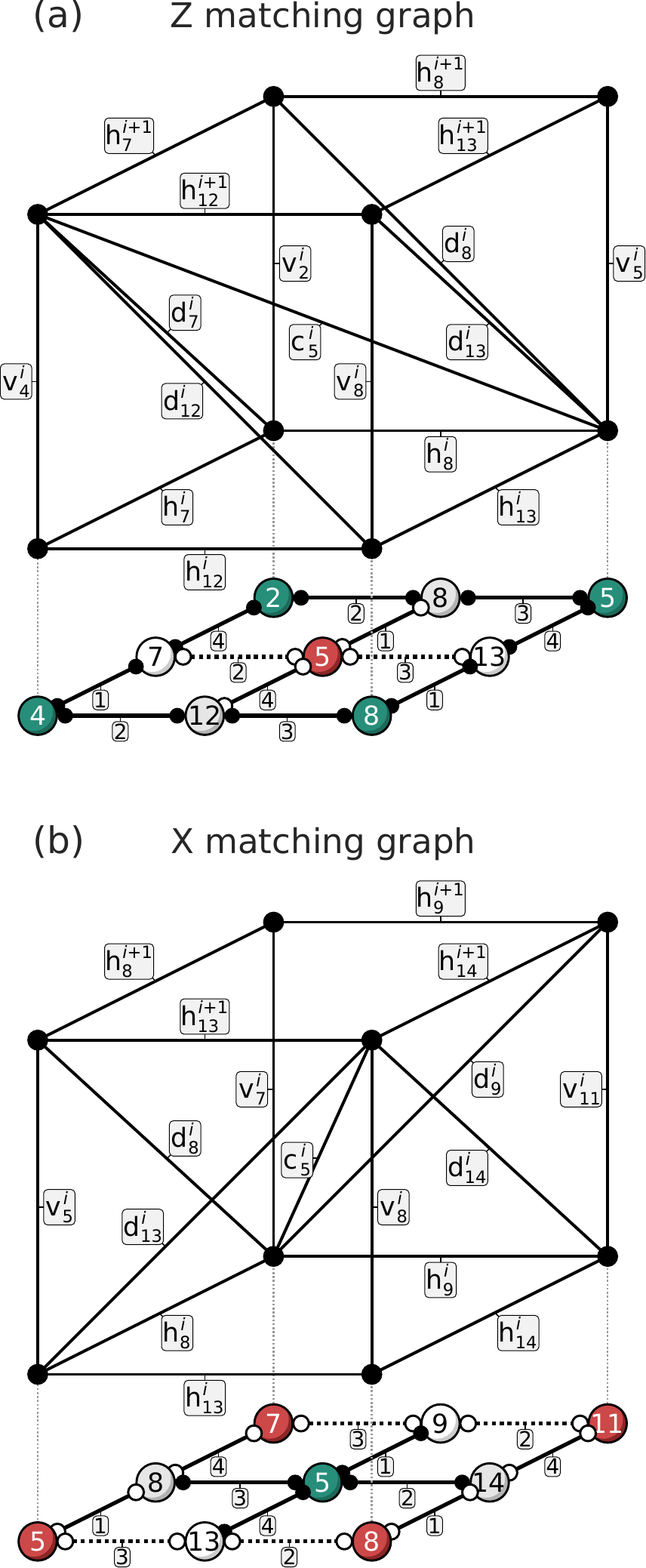}
	\caption{Section of the $Z$ (a) and $X$ (b) matching graphs, here with labels corresponding to the code distance $d=5$ similar to Fig.~\ref{fig:surface}. Each two-mode gate is labeled with the number corresponding to the step in the $i$'th round of syndrome measurements.}
	\label{figC:graphs}
\end{figure}

In this appendix, we describe the matching graphs used in the MWPM decoding. A section of the $Z$ and $X$ matching graphs is shown in Fig.~\ref{figC:graphs}. Each vertex corresponds to a syndrome measurement and is highlighted when the measurement outcome change relative to the previous measurement of the same measure qubit, indicating an error event. The edges correspond to possible errors, and the job of the decoder is to match pairs of highlighted vertices with the most likely error path, which is then translated into corrections of data qubits.

Each horizontal plane of the matching graphs in Fig.~\ref{figC:graphs} corresponds to one round, $i$, of the syndrome measurements in Fig.~\ref{fig:surface}(b,c). We distinguish between 4 different types of edges: horizontal edges, $h$; vertical edges, $v$; diagonal edges, $d$; and cross edges, $c$. In the case where errors only occur in between syndrome measurements (i.e. before and after the syndrome measurement circuits in Fig.~\ref{fig:surface}(b,c)), errors on data qubits correspond to $h$-edges, while errors in the syndrome measurement readout correspond to $v$-edges. These are often the only edges included in the surface code matching graphs when errors during the syndrome measurements are not considered. However, for the surface-4-GKP code with GKP quadrature correction during the syndrome measurements, most qubit errors are induced during the syndrome measurements, requiring the additional $d$- and $c$-edges (often referred to as space-time edges) in the matching graphs for optimal decoding \cite{FowlerEdgeWeights}.

Below we go through all possible qubits errors which may be induced by each GKP quadrature correction in each of the four steps in the syndrome measurements and describe the corresponding edge in the matching graphs. For simplicity, in this work, we consider only uncorrelated single-qubit errors. Since the surface code consists of two-mode gates, two-qubit errors are possible as well, but we assume them to be negligible in the considered squeezing range. The effect of two-qubit errors on the matching graphs is shown in \cite{FowlerEdgeWeights}. Finally, in section \ref{appC:5} we comment on the graph boundaries, edge weighting, and qubit correction.

\subsection{Step 1}
After the first set of two-mode gates in the syndrome measurements (step 1), we perform GKP quadrature correction. At this point, below we consider single-qubit Pauli-$X$ errors, $\hat{X}=e^{\pm i\sqrt{\pi}\hat{p}}$, and Pauli-$Z$ errors, $\hat{Z}=e^{\pm i\sqrt{\pi}\hat{q}}$, (corresponding to $\sqrt{\pi}$ displacement in $\hat{q}$ and $\hat{p}$, respectively) in each data and measure qubit. In all the following we will refer to the $j$'th data, measure-$Z$, and measure-$X$ qubit as $D_j$, $Z_j$, and $X_j$, and we will use the labeling in Fig.~\ref{figC:graphs} for the distance $d=5$ code as an example.

For an odd data qubit, say $D_{13}$, an $\hat{X}$ error is detected by $Z_5$ in step 4 of the current round $i$, while it will be detected by $Z_8$ in step 1 of the following round $i+1$. As a result, this error corresponds to the edge $d_{13}^i$ of the $Z$-graph. A $\hat{Z}$ error will be detected in $X_8$ and $X_5$ in steps 2 and 3 of the current round $i$ and corresponds to $h_{13}^i$ of the $X$-graph.

For an even data qubit, say $D_8$, an $\hat{X}$ error is detected by $Z_2$ and $Z_5$ in steps 2 and 3 of the current round $i$, and corresponds to $h_8^i$ of the $Z$-graph. A $\hat{Z}$ error is detected by $X_7$ in step 4 of the current round $i$, and by $X_5$ in step 1 of the following round $i+1$. Thus this error corresponds to $d_8^i$ of the $X$-graph.

For the measure-$Z$ qubit $Z_5$, an $\hat{X}$ error propagates to $D_{8,13,14}$ through $\hat{C}_Z(1)$ in steps 2, 3, and 4 where they cause a $\hat{Z}$ error, which eventually will highlight $X_7$ and $X_{11}$ in step 4 of this round $i$, indicating a false error on $D_9$ instead of on $D_{8,13,14}$. However, applying a $\hat{Z}$ correction on $D_9$ constitutes together with the $\hat{Z}$ errors on $D_{8,13,14}$ a code stabilizer, and the error is successfully corrected. Thus, an $\hat{X}$ error on $Z_5$ corresponds to $h_9^i$ of the $X$-graph. A
$\hat{Z}$ error on $Z_5$ does not propagate through the $\hat{C}_Z(1)$ gates, and becomes instead a detection error at the end of the syndrome measurement, corresponding to $v_5^i$ of the $Z$-graph.

For the measure-$X$ qubit $X_5$, an $\hat{X}$ error does not propagate through the $\hat{C}_X(\pm1)$ gates, and becomes a detection error at the end of the syndrome measurement, corresponding to $v_5^i$ of the $X$-graph. A $\hat{Z}$ error propagates as $\hat{X}$ errors to $D_{7,12,13}$ through $\hat{C}_X(\pm1)$ gates, which eventually highlights $Z_2$ and $Z_5$ in step 4 of this round, indicating a false error on $D_8$. Applying an $\hat{X}$ correction on $D_8$ constitutes together with the $\hat{X}$ errors on $D_{7,12,13}$ a code stabilizer, and the error is successfully corrected. Thus a $\hat{Z}$ error on $X_5$ corresponds to $h_8^i$ of the $Z$-graph.

\subsection{Step 2}
Below we consider single-qubit errors induced in the GKP quadrature correction after the second set of two-mode gates in the syndrome measurements (step 2).

An $\hat{X}$ error on an odd data qubit is detected similarly to an $\hat{X}$ error induced in step 1. A $\hat{Z}$ error on an odd data qubit, say $D_{13}$, is detected by $X_5$ in step 3 of this round $i$, while detected by $X_8$ in step 2 of round $i+1$. Thus the corresponding edge is $d_{13}^i$ of the $X$-graph.

An $\hat{X}$ error on an even data qubit, say $D_8$, is detected by $Z_5$ in step 3 of this round $i$, while detected by $Z_2$ in step 2 of round $i+1$, and the corresponding edge is $d_8^i$ of the $Z$-graph. A $\hat{Z}$ error on an even data qubit is detected in the same way as a $\hat{Z}$ error induced in step 1.

An $\hat{X}$ error on the measure-$Z$ qubit $Z_5$ propagates through $\hat{C}_Z(1)$ to $D_8$ and $D_{13}$ in steps 3 and 4 as $\hat{Z}$ error. The $\hat{Z}$ error on $D_8$ will be detected by $X_7$ in step 4 of this round $i$, while the $\hat{Z}$ error on $D_{13}$ will be detected by $X_8$ in step 2 of round $i+1$. As a result, the corresponding edge is $c_5^i$ of the $X$-graph. Similar to step 1, a $\hat{Z}$ error corresponds to a detection error, i.e. $v_5^i$ of the $Z$-graph.

An $\hat{X}$ error on measure qubit $X_5$, similar to step 1, corresponds to a detection error, i.e. $v_5^i$ of the $X$-graph. A $\hat{Z}$ error on $X_5$ propagates through $\hat{C}_X(\pm1)$ to $D_{12}$ and $D_{13}$ in steps 3 and 4 as $\hat{X}$ error. The $\hat{X}$ on $D_{13}$ will be detected at $Z_5$ in step 4 of this round $i$, while the $\hat{X}$ error on $D_{12}$ will be detected by $Z_4$ in step 2 of the following round $i+1$. As a result, the corresponding edge is $c_5^i$ of the $Z$-graph.

\subsection{Step 3}
Below we consider single-qubit errors induced in the GKP quadrature correction after the third set of two-mode gates in the syndrome measurements (step 3).

An $\hat{X}$ error on an odd data qubit is detected similarly to an $\hat{X}$ error induced in steps 1 and 2. A $\hat{Z}$ error on an odd data qubit, say $D_{13}$ is detected by $X_8$ and $X_5$ in steps 2 and 3 in the following round $i+1$, and the corresponding edge is $h_{13}^{i+1}$ of the $X$-graph.

An $\hat{X}$ error on an even data qubit, say $D_8$, is detected by $Z_2$ and $Z_5$ in steps 2 and 3 of the following round $i+1$, and the corresponding edge is $h_{8}^{i+1}$ of the $Z$-graph. A $\hat{Z}$ error on an even data qubit is detected in the same way as a $\hat{Z}$ error induced in steps 1 and 2.

An $\hat{X}$ error on measure-$Z$ qubit $Z_5$ propagates through $\hat{C}_Z(1)$ to $D_{13}$ in step 4 as a $\hat{Z}$ error and is detected by $X_8$ and $X_5$ in steps 2 and 3 of the following round $i+1$. Thus the corresponding edge is $h_{13}^{i+1}$ of the $X$-graph. Similar to steps 1 and 2, a $\hat{Z}$ error corresponds to a detection error, i.e. $v_5^i$ of the $Z$-graph.

An $\hat{X}$ error on measure qubit $X_5$, similar to steps 1 and 2, corresponds to a measure, i.e. $v_5^i$ of the $X$-graph. A $\hat{Z}$ error on $X_5$ propagates through $\hat{C}_X(1)$ as an $\hat{X}$ error to $D_{12}$ in step 4, where it will be detected by $Z_4$ and $Z_8$ in steps 2 and 3 of the following round $i+1$. Thus the corresponding edge is $h_{12}^{i+1}$ of the $Z$-graph.

\subsection{Step 4}
Errors on data qubits, $D_j$, induced in the GKP quadrature correction after the final fourth set of two-mode gates in the syndrome measurements (step 4) will only be detected in the following round $i+1$, and the corresponding edges are $h_j^{i+1}$ of the $X$- and $Z$-graph.

Since measure qubits are now to be measured after this fourth step of the syndrome measurements, they do not go through GKP quadrature correction, and we do not induce any qubit error. Measure-$Z$ qubits are measured in the $\hat{p}$ quadrature, and a wrong measurement outcome (caused by finite squeezing of the GKP spikes in the measured quadrature) corresponds to $v_j^{i}$ in the $Z$-graph. Similar, measure-$X$ qubits are measured in the $\hat{q}$ quadrature, and a wrong measurement outcome corresponds to $v_j^{i}$ in the $X$-graph.

\subsection{Graph boundaries, weighting, and correction}\label{appC:5}
As described in section \ref{sec:GKP}, \ref{sec3:code}, and appendix \ref{app:results}, from the residual analogue information in the GKP quadrature correction, we can infer the probability of having induced a qubit error by the GKP quadrature correction, which we then use for weighting of the corresponding above-described edges in the matching graphs. With each edge corresponding to several different errors, for each edge, the probabilities from multiple GKP quadrature corrections are combined into $p_\text{tot}$ by Eq.~\eqref{eq3:Ptot}, and the corresponding edge is finally weighted by $\log_2(p_\text{tot})$.

At the boundary of the code, not all diagonal edges are used. Instead, the corresponding horizontal edge in the following round $i+1$ is used. As an example, consider an $\hat{X}$ error induced on $D_3$ in the distance $d=5$ code (see Fig.~\ref{fig:surface}(a) for labeling). This error will be detected by $Z_2$ in the following round $i+1$, but there are no other measure-$Z$ qubits to detect the error in the current round $i$. Thus, instead of weighting $d_3^{i}$, in this case, the error probability is included in the weight of $h_3^{i+1}$.

In the final round $i=d+1$ where an ideal syndrome measurement is performed, all data qubits are first corrected with perfect GKP quadrature corrections (using $\ket{\varnothing}_\text{GKP}$-states of infinite squeezing) before measuring the stabilizers of the surface code using measure qubits of GKP-states with infinite squeezing. As a result, this final round only requires horizontal edges.

After the MWPM decoding, vertices in the matching graphs are matched with error paths following edges with the most likely errors based on the edge weighting. A vertical edge in an error path, corresponding to a detection error, requires no correction on data qubits. A horizontal or diagonal edge, $h_j^i$ or $d_j^i$, in an error path in the $Z$ or $X$ graph requires an $\hat{X}$ or $\hat{Z}$ correction on data qubit $D_j$, respectively. Finally, a cross edge, $c_j^i$, in an error path in the $Z$ or $X$ graph requires two $\hat{X}$ or $\hat{Z}$ corrections on neighboring data qubits, respectively: As an example, including $c_5^i$ of Fig.~\ref{figC:graphs}(a) in an error path requires $\hat{X}$ corrections on $D_7$ and $D_8$, or on $D_{12}$ and $D_{13}$. Similarly, including $c_5^i$ of Fig.~\ref{figC:graphs}(b) in an error path requires $\hat{Z}$ corrections on $D_8$ and $D_{13}$, or on $D_9$ and $D_{14}$.

\end{document}